\documentclass[graybox]{svmult}
\usepackage{cite}  

\usepackage{type1cm}        
\usepackage{makeidx}         
\usepackage{graphicx}        
\usepackage{multicol}        
\usepackage{slashbox}
\usepackage[bottom]{footmisc}

\usepackage{newtxtext}       %
\usepackage[varvw]{newtxmath}       
\usepackage{mathbbol}

\usepackage{afterpage}
\usepackage{longtable}
\usepackage{amsmath}
\usepackage{cancel}
\usepackage{hyperref}

\makeindex             

\def\NeqOne{{{\cal N} = 1}}

\def\NeqFour{{{\cal N} = 4}}
\def\NeqFive{{{\cal N} = 5}}

\def\NeqEight{{{\cal N} = 8}}

\def\tree{{\rm tree}}

\def\eps{\epsilon}

\def\cN{{\mathcal N}}
\def\n{{\tilde n}}
\def\tn{{\tilde n}}

\def\tJ{\tilde J}

\def\be{\begin{equation}}
\def\ee{\end{equation}}
\def\eea{\end{eqnarray}}
\def\bea{\begin{eqnarray}}
\def\nn{\nonumber}

\def\sect#1{Sect.~{\ref{#1}}}

\def\fig#1{Fig.~{\ref{#1}}}

\def\eqn#1{Eq.~(\ref{#1})}

\def\tab#1{Table~{\ref{#1}}}

\newcommand{\fancyM}{{ \cal K}_{\rm G} \, }

\def\spa#1.#2{\left\langle#1\,#2\right\rangle}
\def\spb#1.#2{\left[#1\,#2\right]}

\def\spash#1.#2{\spa{\smash{#1}}.{\smash{#2}}}
\def\spbsh#1.#2{\spb{\smash{#1}}.{\smash{#2}}}
\def\sand#1.#2.#3{%
\left\langle\smash{#1}{\vphantom1}^{-}\right|{#2}%
\left|\smash{#3}{\vphantom1}^{-}\right\rangle}
\def\sandpp#1.#2.#3{%
\left\langle\smash{#1}{\vphantom1}^{+}\right|{#2}%
\left|\smash{#3}{\vphantom1}^{+}\right\rangle}
\def\sandpm#1.#2.#3{%
\left\langle\smash{#1}{\vphantom1}^{+}\right|{#2}%
\left|\smash{#3}{\vphantom1}^{-}\right\rangle}
\def\sandmp#1.#2.#3{%
\left\langle\smash{#1}{\vphantom1}^{-}\right|{#2}%
\left|\smash{#3}{\vphantom1}^{+}\right\rangle}
\def\sand#1.#2.#3{%
\left\langle\smash{#1}{\vphantom1}\right|{#2}%
\left|\smash{#3}{\vphantom1}\right\rangle}
\def\sandp#1.#2.#3{%
\left\langle\smash{#1}{\vphantom1}^{-}\right|{#2}%
\left|\smash{#3}{\vphantom1}^{+}\right\rangle}
\def\sandpp#1.#2.#3{%
\left\langle\smash{#1}{\vphantom1}^{+}\right|{#2}%
\left|\smash{#3}{\vphantom1}^{+}\right\rangle}
\def\sandmm#1.#2.#3{%
\left\langle\smash{#1}{\vphantom1}^{-}\right|{#2}%
\left|\smash{#3}{\vphantom1}^{-}\right\rangle}
\def\sandpm#1.#2.#3{%
\left\langle\smash{#1}{\vphantom1}^{+}\right|{#2}%
\left|\smash{#3}{\vphantom1}^{-}\right\rangle}
\def\sandmp#1.#2.#3{%
\left\langle\smash{#1}{\vphantom1}^{-}\right|{#2}%
\left|\smash{#3}{\vphantom1}^{+}\right\rangle}

\def\calMthree{{\cal M}^{\tree}_3}
\def\calMfour{{\cal M}^{\tree}_4}
\def\calMfive{{\cal M}^{\tree}_5}

\DeclareMathAlphabet\mathbfcal{OMS}{cmsy}{b}{n}

\newbox\charbox
\newbox\slabox
\def\s#1{{
        \setbox\charbox=\hbox{$#1$}
        \setbox\slabox=\hbox{$/$}
        \dimen\charbox=\ht\slabox
        \advance\dimen\charbox by -\dp\slabox
        \advance\dimen\charbox by -\ht\charbox
        \advance\dimen\charbox by \dp\charbox
        \divide\dimen\charbox by 2
        \raise-\dimen\charbox\hbox to \wd\charbox{\hss/\hss}
        \llap{$#1$} }}

\begin{document}

\title*{Supergravity amplitudes, the double copy and ultraviolet behavior\thanks{Invited chapter for the {\it Handbook of Quantum Gravity}~~(Eds. C.~Bambi, L.~Modesto, and I.~L.~Shapiro, Springer 2023).}}


\author{Z.~Bern, J.~J.~Carrasco, M.~Chiodaroli, H.~Johansson and R.~Roiban}
\institute{Zvi Bern \at Mani L. Bhaumik Institute for Theoretical Physics, UCLA, Los Angeles, CA 90024, USA.
      \email{bern@physics.ucla.edu} 
\and John Joseph Carrasco \at Northwestern University, Evanston, IL 60208, USA.
   \email{carrasco@northwestern.edu} 
\and   Marco Chiodaroli \at  Department of Physics and Astronomy,  Uppsala University, Box 516, 75120 Uppsala, Sweden.
\email{marco.chiodaroli@physics.uu.se}
\and Henrik Johansson \at Department of Physics and Astronomy,  Uppsala University, Box 516, 75120 Uppsala, Sweden \\
                     and Nordita, Stockholm University and KTH Royal Institute of Technology,
                      Hannes Alfv\'{e}ns v\"{a}g 12, 10691 Stockholm, Sweden. \email{henrik.johansson@physics.uu.se}
\and Radu Roiban \at Institute for Gravitation and the Cosmos,
                          Pennsylvania State University, University Park, PA 16802, USA. \email{radu@phys.psu.edu}
}

%
%

\maketitle

${}$
\vspace{-1.5truecm}

\abstract*{
In this chapter, we present a scattering-amplitudes perspective on supergravity, and describe its application to the study of ultraviolet properties of supergravity theories at high loop orders. The basic on-shell tools that make such calculations feasible are reviewed, including generalized unitarity, color-kinematics duality, and the double-copy construction.  We also outline the web of theories connected  by the double copy.  The results of various calculations of potential ultraviolet divergences are summarized.  These include puzzling {enhanced ultraviolet cancellations}, for which no symmetry-based understanding currently exists, showing that there is much more to learn about the ultraviolet properties of supergravity theories.      We finally comment on future calculations that should help resolve the puzzles.
}

\abstract{
In this chapter, we present a scattering-amplitudes perspective on supergravity, and describe its application to the study of ultraviolet properties of supergravity theories at high loop orders. The basic on-shell tools that make such calculations feasible are reviewed, including generalized unitarity, color-kinematics duality, and the double-copy construction.  We also outline the web of theories connected  by the double copy.  The results of various calculations of potential ultraviolet divergences are summarized.  These include puzzling {enhanced ultraviolet cancellations}, for which no symmetry-based understanding currently exists, showing that there is much more to learn about the ultraviolet properties of supergravity theories.      We finally comment on future calculations that should help resolve the puzzles.
}


\section{Introduction}
\label{IntroSection}

Since the discovery of supergravity~\cite{Freedman:1976xh, Deser:1976eh}, fully understanding its hidden symmetries, basic structures and allowed generalizations have been highly nontrivial challenges. A surprising  hidden structure, shared by many supergravity theories, is the double-copy~\cite{Kawai:1985xq, BCJ, BCJLoop, Bern:2019prr} which, in its simplest form, implies that gravity scattering amplitudes can be obtained directly from gauge-theory quantities. When combined with the unitarity method~\cite{UnitarityMethod, Fusing, FiveLoop}, the double copy has made possible multi-loop calculations needed to explicitly determine the ultraviolet properties of various supergravity theories.

Power counting arguments following from the dimensionful nature of Newton’s constant suggest that all point-like theories of gravity should be expected to be ultraviolet divergent at some sufficiently high loop order.  Conversely, the absence of such divergences would hint at the existence of some novel structure or symmetry responsible for its cancellation.  
Therefore, studies of ultraviolet divergences can shed light on basic structures and symmetries present in particular supergravities, exposing properties that are otherwise hidden.
Traditional off-shell superspace methods become challenging beyond one loop.
The double copy and the unitarity method provided a means to explore multiloop questions in pure supergravities, and have been used for such studies at three, four and five loops~\cite{Bern:2007hh, Bern:2009kd, SimplifyingBCJ, Bern:2012gh, Bern:2013qca, Bern:2013uka, Bern:2014sna, UVFiveLoops}. 
In this chapter, we will explain these ideas and emphasize puzzling {\it enhanced ultraviolet cancellations}.  We will also briefly describe a web of theories linked via the double copy.

Arguments based on supersymmetry and known duality symmetries reveal that ultraviolet divergences are delayed to surprisingly high-loop orders. In particular, such arguments show that $\NeqEight$ supergravity is ultraviolet finite through at least six loops~\cite{Bern:2007hh, Bern:2009kd, SimplifyingBCJ,  Green:2010sp, Bjornsson:2010wm, Bossard:2010bd, Beisert2010jx, Bossard:2011tq}. 
From this perspective, a natural question is then whether all supergravity theories must necessarily diverge in the ultraviolet at some loop order, or whether a hidden structure prevents the appearance of divergences at least for certain special theories.

Studies of unitarity cuts in $D=4$ suggest the interesting possibility that divergences in ${\cal N}=8$ supergravity may be further delayed~\cite{Herrmann:2018dja,Edison:2019ovj} in this dimension.  
So far, the only explicitly known divergence in a pure supergravity arises in $\NeqFour$ supergravity at four loops~\cite{Bern:2013uka}. The interpretation of this divergence is, however, complicated by the presence of a $U(1)$ duality anomaly which might be behind its appearance~\cite{Bern:2017rjw, Bern:2019isl}.
Such anomalies are absent in ${\cal N} \ge 5$ supergravities~\cite{Freedman:2017zgq}, implying that it is best to focus on these theories. 
A remarkable case is that of $\NeqFive$ supergravity: it does not diverge at four points at the four-loop order~\cite{Bern:2014sna} despite there being no known symmetry mechanism protecting it~\cite{Bossard:2011tq, Freedman:2018mrv}, and therefore provides the only known four-dimensional example of an enhanced cancellation.  
It is therefore crucial to determine whether the four-point amplitude in $\NeqFive$ supergravity is ultraviolet-finite at five loops.
A positive outcome, perhaps supported by the identification of an additional 
symmetry or structure that can protect against ultraviolet divergences, 
may point towards the possibility of the all-order finiteness of this theory and ultimately inspire a proof. This would be along the same lines as the proofs of ultraviolet finiteness~\cite{MandelstamN4SYM, BrinkN4SYM, HoweStellN4SYM} of $\NeqFour$ SYM theory~\cite{Brink:1976bc, Gliozzi:1976qd} that followed explicit calculations through three loops~\cite{Jones:1977zr, Poggio:1977ma, Grisaru:1979wc,Grisaru:1980nk}. 
Such an all-orders proof in a supergravity theory would be highly nontrivial because of the still mysterious nature of enhanced cancellations.
The existence of an undiscovered symmetry or structure would likely have a fundamental impact on our understanding of quantum gravity. 

The study of the ultraviolet properties of theories of gravity has a long history, starting with the seminal work of 't~Hooft and Veltman~\cite{tHooft:1974toh}, who showed that pure Einstein gravity is finite at one loop but divergent in the presence of matter, which is also confirmed in other examples~\cite{Deser:1974xq, Fischler:1979yk}. 
Subsequently, Goroff and Sagnotti showed that pure Einstein gravity diverges at two loops~\cite{Goroff:1985th, vandeVen:1991gw, Bern:2015xsa}. The ultraviolet behavior improves with the addition of supersymmetry. By the late 1970s it was known that pure ungauged supergravities cannot have divergences prior to three loops~\cite{Grisaru:1976nn, Tomboulis:1977wd, Deser:1978br}. 
The consensus reached from studies in the 1980s was that all pure supergravity theories would likely diverge at the third loop order (see, for example, Ref.~\cite{Green:1982sw, Marcus:1984ei, Howe:1988qz}), though by making additional assumptions one can raise the loop order of the predicted potential divergences~\cite{Grisaru:1982zh}.

The duality between color and kinematics and the associated double-copy offer a perspective on gravitational interactions that is radically different from the conventional geometric one.
It states that, after a suitable rearrangement of the contributing diagrams,  there is a simple procedure to convert gauge-theory scattering amplitudes to gravitational ones via a replacement of color factors by corresponding kinematic factors.
It may very well be true that scattering amplitudes in all (super)gravity theories can be written in a double-copy form in terms of gauge theory~\cite{Chiodaroli2015wal, Anastasiou:2017nsz}. For the purpose of this review we will focus on the simplest cases of ungauged ${\cal N} \ge 4$ supergravity theories, which have particularly simple double-copy constructions. 
More generally, the double copy can be applied to non-gauge theories with some Lie-algebra symmetry, leading to a veritable web of interrelated theories~\cite{Cheung:2017ems, Bern:2019prr}. Remarkably, the double copy can also be used to relate classical solutions of gravity to those of gauge theory; while no general construction exists, many examples are available -- see for example Refs.~\cite{Monteiro2014cda, Luna2015paa} and Ref.~\cite{Bern:2019prr} for a review. 

Explicit calculations carried out using these methods have greatly informed and guided our understanding of ultraviolet properties of supergravity theories. For maximally supersymmetric supergravity, which in $D=4$ is $\NeqEight$ supergravity~\cite{Cremmer:1979up},  calculations conclusively demonstrate that the three-loop four-point amplitudes is finite for space-time dimensions $D<6$~\cite{Bern:2007hh,Bern:2008pv} and at four loops for $D<11/2$~\cite{Bern:2009kd}. In $D=4$, these ultraviolet cancellations were subsequently understood to follow from supersymmetry and the $E_{7(7)}$ duality symmetry of $\NeqEight$ supergravity~\cite{Green:2010sp, Bossard:2010bd, Beisert2010jx, Bossard:2011tq}. A purely supersymmetric explanation was given 
in Ref.~\cite{Bjornsson:2010wm} based on the Berkovits pure-spinor formalism~\cite{Berkovits:2000fe}.  The consensus at the time of this writing is that in $D=4$ a $D^8 R^4$ counterterm is not forbidden by any of the known symmetries, leading to the expectations of a seven-loop divergence.  By analytically continuing to $D = 24/5$, this counterterm corresponds to a five-loop divergence~\cite{Bjornsson:2010wm}.
Such a procedure  does not however account for special $D=4$ properties~\cite{Herrmann:2018dja,Edison:2019ovj} of these amplitudes of this theory.

The most interesting cases exhibiting enhanced ultraviolet cancellations, with no currently known symmetry explanation, include
\begin{enumerate}
    \item pure half-maximal supergravity at two loops in $D=5$~\cite{Bern:2012gh, Tourkine:2012ip, Bossard:2013rza, Bern:2013qca}; 
    \item  pure $\NeqFour$ supergravity at three loops in $D=4$~\cite{Bern:2007hh, Bern:2008pv, Bossard:2011tq, Tourkine:2012ip};
    \item  $\NeqFive$ supergravity at four loop in $D=4$~\cite{Bern:2014sna, Bossard:2011tq}.
\end{enumerate}
As already emphasized, the last example is perhaps the most appealing one because the relevant cancellations occur in $D=4$ and the theory does not suffer from a $U(1)$ anomaly~\cite{Freedman:2017zgq} present in $\NeqFour$ supergravity~\cite{MarcusAnomaly, Carrasco:2013ypa}. The absence of this anomaly and the known presence of enhanced cancellation makes it a candidate ultraviolet-finite theory; verifying this property at the next (five) loop order is within the reach of currently available methods.

Additional cancellations found in the unitarity cuts of $\NeqEight$ supergravity~\cite{Herrmann:2018dja, Edison:2019ovj}  suggest that this theory may very well also display enhanced cancellation at seven loops in $D=4$, rendering it ultraviolet-finite to at least this loop order. A direct test of this expectation will require further technical advances, given the difficulty of carrying out seven loop calculations.

What might be the origin of the observed enhanced ultraviolet cancellations?  While we do not have a complete picture, in the case of half-maximal supergravity at two loops in $D=5$, an  explanation is revealed by the duality between color and kinematics~\cite{Bern:2012gh}. 
Together with the double copy, it relates the two-loop enhanced supergravity cancellations to the cancellation of potential divergences with forbidden color factors in the single-copy gauge theories.
This case is particularly simple at two-loop order because the supergravity amplitudes are linear combinations of gauge-theory amplitudes even after the loop integration is carried out.  This is connected to the rather simple structure of the four-point two-loop integrand in maximally supersymmetric Yang--Mills theory.
A similar analysis at three and higher loops is much more difficult because the simple relation between {\em integrated} gauge and gravity amplitudes no longer holds~\cite{Bern:2017lpv}. 
Other proposals  for possible symmetry-based explanations of enhanced cancellation have been put forth in Refs.~\cite{Kallosh:2018wzz,Gunaydin:2018kdz}. New results on generalized symmetries provide new mechanisms which forbid the appearance of certain operators in the effective action~\cite{Gaiotto:2014kfa,Komargodski:2020mxz,Cordova:2022ruw}. 
It would be important to understand whether generalized symmetries can shed light on the mysterious enhanced cancellation observed in supergravity theories.

In \sect{OnshellSection} of this chapter, we start by first reviewing tree-level amplitudes and the associated on-shell superspaces convenient for describing scattering amplitudes in supersymmetric gauge and gravity theories. This section also explains the duality between color and kinematics and the double copy at tree level.  
The unitarity method, producing loop-level amplitudes starting from tree amplitudes, and some of its consequences are outlined in \sect{OnshellSection}. 
In \sect{LoopLevelSection} we describe the loop-level methods that are used to explore ultraviolet properties that are then summarized in \sect{UVSection}.   
The web of theories, linking various gravitational and non-gravitational theories, is summarized in \sect{WebSection}. 
Finally, in \sect{ConclusionSection} we present our conclusions together with an outlook on the application of scattering amplitude methods to supergravity theories.


\section{Tree-level amplitudes and their properties}
\label{OnshellSection}

There are two complementary advances that have made it possible to climb multiloop orders in supergravity theories starting from tree-level amplitudes.  The first is the realization that the data encoded in tree-amplitudes completely constrains loop integrands.  In this section we describe tree-level amplitudes which then feed into loop level via on-shell unitarity methods described in \sect{LoopLevelSection}.  The second advance follows from the realization that calculations in perturbative gravity do not need be more complicated than in gauge theory. This is reflected in the double-copy construction that relates gravity amplitudes to Yang--Mills quantities, or in the supersymmetric context,  relates supergravity amplitudes to super-Yang--Mills (SYM) quantities. We can encode the spectrum of supergravity theories in terms of the spectrum of two copies of SYM theories.  As much of the book-keeping in scattering calculation involves tracking particle states, we will spend the first part of this tree-level section describing on-shell superspace in gauge theory, discuss double-copy constructions, and then clarify how this encodes supergravity states. 

 In supersymmetric theories scattering states can be conveniently stratified into so-called on-shell superfields.  These superfields realize  linearized  supersymmetry. In contrasted with off-shell Green's functions which exhibit both the linear and nonlinear part of symmetries, scattering amplitudes require only linearized supersymmetry, with non-linear terms projected out by the LSZ reduction. 
With such an organization of the asymptotic states, it is natural to collect amplitudes 
into superamplitudes which, similarly with the on-shell superfields, manifest linearized supersymmetry.

Several on-shell superspaces have been put forth; similarly to their off-shell counterpart, their formulation is dimension-dependent. Their power however relies 
on the existence of unconstrained Grassmann variables which form a fundamental representation of the R-symmetry group\footnote{Such superspaces are typically referred to as ``chiral".} in terms of which superfields are 
unconstrained polynomials whose coefficients are the physical asymptotic states.
Such Grassmann variables were originally introduced in Ref.~\cite{Ferber:1977qx} to 
yield a supersymmetric extension of twistors and subsequently used in Ref.~\cite{Nair}
in the context described here.

In this section we discuss the four-dimensional and six-dimensional maximally-supersymmetric superspaces and the corresponding tree-level superamplitudes, from which lower-supersymmetric cases can be obtained by truncation, tree-level color-kinematics duality and the double copy and some of their consequences.

\vskip -.7 cm $\null$
\subsection{Tree-level superamplitudes in various dimensions}
\label{SuperTreeSubsection}
\vskip -.2 cm 

In the context of the double copy, it is convenient to encode supergravity states in terms of the states of supersymmetric gauge theories.  We begin with maximal supersymmetry.  The R-symmetry of the maximally-supersymmetric gauge theory in four dimensions is $SU(4)$, so the on-shell superspace is constructed in terms of four set of Grassmann variables $\eta^a$, with $a=1 \cdots 4$ transforming in its fundamental representation. 
All physical states of this theory are labeled by their helicity and form the vector multiplet of the ${\cal N}=4$ superalgebra.  These can be combined into the CPT-self-conjugate on-shell superfield,
\begin{equation}
\Phi(\eta)=g^++\eta^af^+_a+\frac{1}{2}\eta^a\eta^b\phi_{ab}+\frac{1}{3!}\epsilon_{abcd}\eta^a\eta^b\eta^cf^{d-}+\frac{1}{4!}\epsilon_{abcd}\eta^a\eta^b\eta^c\eta^dg^- . \hskip .2 cm 
\label{Neq4multiplet}
\end{equation}   
Similarly, the states of the maximally-supersymmetric supergravity theory in four dimensions, ${\cal N}=8$ supergravity, can be organized in a CPT-self-conjugate multiplet which is now build out of eight Grassmann variables, $\eta^A, A=1\dots 8$, it exhibits $SU(8)$ R symmetry and contains states up to helicity $\pm 2$ ---as we will see we can encode the $\mathcal{N}=8$ supergravity states in terms of an outer product of the $\mathcal{N}=4$ SYM states.  Similarly with off-shell superspaces, supersymmetry transformations relating the component fields are realized as shifts $\eta^a\mapsto \eta^a+\epsilon^a$ of the Grassmann variables. 
The simplest amplitudes are the tree-level so called maximally-helicity-violating  (MHV) $n$-point color-ordered amplitudes.  The MHV amplitudes are the nonvanishing ones with the maximum imbalance between positive and negative helicity states. As standard for gauge-theory tree-level scattering amplitudes we also use a color-ordered format which effectively strips the color factors from the amplitudes. (See Ref.~\cite{ManganoParkeReview, TasiLance}) for details on color ordering.) With the above organization of asymptotic states, the tree-level MHV $n$-point color-ordered amplitudes of maximally-supersymmetric gauge theory can be packaged into the $n$-point color-ordered superamplitude~\cite{Witten:2003nn},
\begin{equation}
A^{\rm MHV}_n(1,2,{\cdots},n)=\frac{i}{\prod_{j=1}^n\langle j \, (j+1)\rangle}\,
\delta^{(8)}\big(Q^{a\alpha} \big)\,,
\label{MHVSuperAmplitudeSYM}
\end{equation}   
where leg $n+1$ is identified with leg $1$, 
$
Q^{a\alpha} \equiv \sum_{j=1}^n\lambda^{\alpha}_j\eta_j^a
$
is the total supermomentum, and the delta function imposing its conservation may be thought of as a superpartner of the momentum conservation delta function.
In the definition of $Q^{a\alpha}$, $\lambda_i^\alpha$ and its conjugate ${\bar\lambda}_i^{\dot\alpha}$ are spinors solving the massless
mass shell condition for external lines,
$
p_i^\mu {\bar\sigma}_\mu^{\alpha\dot\alpha} =  \lambda_i^\alpha{\bar\lambda}_i^{\dot\alpha}
$.
The Grassmann delta function can be rewritten as
\begin{equation}
\delta^{(8)}\big(Q^{a\alpha} \big) =
\delta^{(8)}\Biggl(\sum_{j=1}^n\lambda^{\alpha}_j\eta_j^a\Biggr)
= \prod_{a=1}^4\sum_{i<j}^n\langle ij\rangle \, \eta_i^a\eta_j^a\,.
\label{DeltaSpinor}
\end{equation}
Component amplitude are extracted either by multiplication with the desired states written as superfields and integration over all the Grassmann parameters for all external states~\cite{Witten:2003nn, ArkaniHamed2008gz}, or by selecting from Eq.~\eqref{MHVSuperAmplitudeSYM} the terms with the desired monomials in the  Grassmann parameter, see e.g. Ref.~\cite{ElvangFreedmanN8superspace, ArkaniHamed2008gz}.

We note that the coefficient of the Grassmann delta function in Eq.~\eqref{MHVSuperAmplitudeSYM} can be conveniently written in terms of the  tree-level MHV color-ordered gluon amplitude, $\mathbb{A}^\tree_{n}(1^-,2^-,3^+,\cdots,n^+)$, 
of nonsupersymmetric Yang--Mills theory:
\begin{equation}
A^{\rm MHV}_n(1,2,{\cdots},n)= \frac{\mathbb{A}^\tree_{n}(1^-,2^-,3^+,\cdots,n^+)}{\langle 12\rangle^4}\,
\delta^{(8)}\big(Q^{a\alpha} \big)\,.
\label{MHVsYM}
\end{equation}

The MHV superamplitude of $\mathcal{N}=8$ supergravity has an similarly close  
relationship,
with the tree-level MHV graviton amplitude, $\mathbb{M}^\tree_{n}(1^-,2^-,3^+,\cdots,n^+)$, of Einstein's general relativity,
\begin{equation}
 \label{MHVsugra}
{\cal M}_n^{\rm MHV}=\frac{\mathbb{M}^\tree_{n}(1^-,2^-,3^+,\cdots,n^+)}{\langle 12\rangle^8}\,
\delta^{(16)} (Q^{a\alpha} )\,.
 \end{equation}
%

On-shell multiplets with reduced supersymmetry can be obtained by suitably truncating the maximally-supersymmetric multiplets. One simply sets to zero the fields that do not belong to the desired ${\cal N}$-extended multiplet. In the remaining terms, certain Grassmann variables will, as a result of this truncation, always appear together. They transform under the complement of $SU({\cal N})$ inside the maximal R symmetry.
The corresponding MHV superamplitudes are  obtained from the maximally supersymmetric MHV amplitude by retaining  the maximum number of Grassmann parameters that do not transform under the reduced supersymmetry in the supermomentum-conservation delta function.
The simplest example is the amplitudes of pure non-supersymmetric Yang--Mills theory; they are just the pure gluon amplitudes of $\NeqFour$ SYM theory, since no superpartners appear in these amplitudes.\footnote{This projection may be interpreted as the restriction to the invariants under the action of a discrete subgroup of R symmetry, and is usually referred to as a ``field-theory orbifold". More involved realizations combine the action of a discrete subgroup of R symmetry with the action of the same subgroup of the gauge group. We refer the reader to Ref.~\cite{Chiodaroli2013upa} a discussion of the effect of such field-theory orbifolds on amplitudes.}

A slightly more involved example is that of the color-ordered MHV tree amplitudes for the minimal gauge multiplets of ${\cal N}<4$ SYM theory. They were shown in~Ref.~\cite{Bern:2009xq} to be given by
\begin{equation}
A^{\rm MHV}_n(1, 2, \ldots, n)=
\frac{ \prod_{a=1}^{\cal N}\delta^{(2)}( Q^a) }
{\prod_{j=1}^n\langle j ~(j+1)\rangle}
\,\,
\Biggl(\sum_{i<j}^n \spa{i}.{j}^{4-{\cal N}} 
\prod_{b={\cal N}+1}^{4} \eta_i^b  \eta_j^b\Biggr) \,.
\label{MHVLessSuperAmplitude}
\end{equation}
The remaining delta function enforces the conservation of the ${\cal N}$ supercharges while the factor in parenthesis comes from keeping the maximum number of Grassmann parameters that do not transform under the $SU({\cal N})$ R symmetry.
See Refs.~\cite{Bern:2009xq,Elvang:2011fx} for a more in-depth discussion of ${\cal
  N}<4$ on-shell superfields and superamplitudes.

With MHV superamplitudes as input, N$^m$MHV superamplitudes---that is superamplitudes which are related by supersymmetry to gluon amplitudes with $m+2$ gluons of negative helicity---can be efficiently constructed either via supersymmetric CSW rules \cite{Elvang:2008vz} or  BCFW recursion relations~\cite{ArkaniHamed2008gz}. 
While extremely efficient analytically, both approaches yield expressions exhibiting spurious poles, which makes them difficult to use for the construction of loop amplitudes via the unitarity method which we discuss in Sect.~\ref{LoopLevelSection}. 
This complication can be avoided by matching them against an ansatz for the amplitude that has only physical poles. 

Higher-loop calculations require regularization and in Section~\ref{sec:regularization} we will review some of its important aspects. 
With dimensional regularization being the preferred method in both non-supersymmetric and component-based supersymmetric calculation  and to set the stage, we briefly discuss the $D=6$ on-shell superspace, constructed in Ref.~\cite{Dennen:2009vk}, and the corresponding superamplitudes, which can be used for this purpose.  

The six-dimensional massless superspace builds on the six-dimensional spinor-helicity formalism of Ref.~\cite{Cheung:2009dc}.\footnote{Massive on-shell superspaces can also be constructed, see e.g. \cite{Arkani-Hamed:2017jhn} for four dimensions (see also \cite{Craig:2011ws, Kiermaier:2011cr,Conde:2016vxs,Conde:2016izb,Ochirov:2018uyq,Herderschee:2019ofc, 
Johansson:2019dnu,Chiodaroli:2021eug}) and \cite{Chiodaroli:2022ssi} for five dimensions, but they are outside the scope of this review.}
Six-dimensional momenta may be written as $4\times 4$ matrices by contracting them with the six-dimensional chiral Dirac matrices. Then, the on-shell condition demands that this matrix has reduced rank. To obtain the five independent degrees of freedom, they are therefore written as
\bea
p^{AB} = \epsilon^{ab} \lambda^A_a\lambda^B_b \,, \hskip 1.1 cm
p_{AB}=\frac{1}{2}\epsilon_{ABCD}p^{CD} = {\tilde\lambda}_A^{\dot a}{\tilde\lambda}_B^{\dot b}
\epsilon_{{\dot a}{\dot b}} \, ,
\eea
where the internal $SU(2)$ symmetry acting on the lower-case indices is identified with the massless little-group symmetry. Note that, unlike the fur-dimensional case, the spinors $\lambda$ and ${\tilde \lambda}$ are nontrivially related.

The choice of Grassmann variables for the construction of superfields relies on two observations: (1) four- and six-dimensional states are related by simple dimensional reduction and (2) the R symmetry of the four-dimensional theory is not manifestly realized by dimensional reduction while the six-dimensional R symmetry is manifest.
It is then perhaps natural to expect that the six-dimensional $(1,1)$ 
on-shell superspace should reduce to four-dimensional ${\cal N}=4$ one 
in which two of the four Grassmann variables are traded for their Fourier-conjugates. The corresponding six-dimensional maximally-supersymmetric vector superfield is
\begin{align}
\Phi^{D=6}(\eta, {\tilde\eta}) & = \phi+\chi^a\eta_a+{\tilde\chi}_{\dot a}{\tilde\eta}^{\dot a}
+\phi'(\eta)^2+g^a_{\ \, \dot a}\eta_a{\tilde\eta}^{\dot a}
\nonumber \\
& \hskip .6 cm \null
+\phi''({\tilde\eta})^2 + 
{\tilde\lambda}_{\dot a}{\tilde\eta}^{\dot a}(\eta)^2 +{\lambda}^{a}{\eta}_{a}({\tilde \eta})^2
+\phi'''(\eta)^2({\tilde \eta})^2 \, .
\label{6dsuperfield}
\end{align}
Two four-dimensional scalars have been absorbed in the $2\times 2$ matrix $g^a_{\ \,\dot a}$ which contains the physical degrees of freedom of the six-dimensional gluon, leaving four physical scalars, $\phi,\dots,\phi'''$.
With this choice of Grassmann variables, the single-particle supermomenta  are
\bea
q_i^A = \lambda^A_{i,a}\eta_{+,i}^a\,,
\hskip 1.1 cm 
{\tilde q}_{iA} = {\tilde \lambda}_{i,A}^{\dot a}{\tilde \eta}_{+,i,{\dot a}} \, .~~
\eea
The four-point gauge-theory superamplitude is \cite{Dennen:2009vk}
\begin{equation}
A_4^\text{tree}(1,2,3,4)=\frac{1}{st} \, \delta^6\biggl(\sum_{i=1}^4 p_i\biggr) \,
\delta^4\biggl(\sum_{i=1}^4 q_i^A \biggr) \,
\delta^4\biggl(\sum_{i=1}^4 {\tilde q}_i^A \biggr) \, .
\label{fourpt6d}
\end{equation}
For the three-point and the higher-point superamplitudes, which are substantially more involved, we refer the reader to the original literature~\cite{Dennen:2009vk, Bern:2010qa}.

As in the four-dimensional superspace, to extract component amplitudes one multiplies Eq.~(\ref{fourpt6d}) by the desired superfields (each containing a single nonzero component field) and integrates over all Grassmann variables.

More generally, on-shell superspace provides a convenient way to carry out sums over physical states which are needed when constructing higher-point tree-level superamplitudes through either the super-BCFW on-shell recursion relation~\cite{ArkaniHamed2008gz, Drummond:2008vq, Drummond:2008cr} or through the super-MHV vertex rules~\cite{GGK,ElvangFreedmanN8superspace,Kiermaier:2009yu, FreedmanUnitarity, Bern:2009xq}  and in loop calculations carried out through the generalized unitarity method that will discuss this in Section~\ref{LoopLevelSection}. 
The main observation is that integration over Grassmann variables acts as identity operator
in the space of states, i.e. in $D=4$
\begin{equation}
    \int d^4\eta \, \Phi_1(\eta)\Phi_2(\eta) = 
    g_1^+g_2^-
    +f^+_{1,a}f^{-a}_{2}
    +\frac{1}{2}\phi_{1ab}\phi_2^{ab}
    +f^{-a}_{1}f^{+}_{2a}
    +g_1^-g_2^+ \,,
    \label{Grassmass}
\end{equation}
and similarly in $D=6$ on-shell superspace. The factor of $1/2$ accounts for the fact that $\phi_{ab}$ is antisymmetric. 

The sum over all the states that can be present on a common external line $m$ of two color-ordered superamplitudes can be written as a Grassmann integral,
\begin{equation}
  \int d^4 \eta_m \, A(1, \dots, m-1, m)\, A(m, m+1 , \dots, n) \,.
\label{Supersum}
\end{equation}
As we will see in later sections, this observation has been used to great effect in $D=4$ and $D=6$ to carry out high-order loop calculations by sewing together tree amplitudes using superspace versions of unitarity.
Similar Grassmann integrals realize the sum over states in superspace versions of BCFW recursions \cite{ArkaniHamed2008gz} or MHV vertex rules~\cite{Kiermaier:2009yu}. 

\vskip -.7 cm $\null$
\subsection{Color-kinematics duality and double copy}
\label{cktrees}
\vskip -.2 cm 

Among the many interesting properties of tree-level gauge-theory scattering amplitudes, color-kinematics duality \cite{BCJ, BCJLoop} stands out by providing a bridge between gauge and gravitational theories. In short, it states that in general dimensions, with a suitable  reorganization of gauge amplitudes in terms of cubic graphs, the kinematic numerator factors have the same algebraic properties as the corresponding color factors. 
The existence of such a reorganization is nontrivial. It has been proven at tree level~\cite{BjerrumMomKernel, MafraExplicitBCJNumerators} from amplitudes perspective. While one might expect that there should exist a Lagrangian understanding of this property, only partial results are  available~\cite{Square, WeinzierlBCJLagrangian, Vaman:2014iwa, Mastrolia:2015maa}.
Analogous constructions of loop-level integrands are available on a case-by-case basis~(see e.g. Refs.~\cite{BCJLoop, SimplifyingBCJ,  Boels:2012ew, Carrasco:2012ca,Bjerrum-Bohr:2013iza, Bern:2013yya, Chiodaroli2013upa, Chiodaroli2014xia, Badger:2015lda, Mafra:2015mja, He:2015wgf, Mogull:2015adi,Chiodaroli:2015rdg, Yang:2016ear, Chiodaroli:2017ngp, Chiodaroli:2017ehv, He:2017spx, Johansson:2017bfl, Chiodaroli:2018dbu, Carrasco:2020ywq, Lin:2021kht, Li:2022tir, Edison:2022smn, Edison:2022jln}. 
The duality between color and kinematics was initially identified~\cite{BCJ} as a property of gauge-theory amplitudes, and was subsequently extended to gauge theories with matter in various representations of the gauge group~\cite{Chiodaroli2013upa, Johansson2014zca, Johansson:2015oia, Carrasco:2020ywq}
and to field theories without gauge fields such as biadjoint $\phi^3$~\cite{Du2011js, OConnellAlgebras, White2016jzc}
and the nonlinear sigma model (NLSM)~\cite{Chen2014dfa,  Du2016tbc, Chen2016zwe,Cheung:2017ems, Carrasco:2019qwr, Pavao:2022kog}. 
Perhaps the most remarkable consequence of color-kinematics duality is that it gives, through the double copy, a means to combining amplitudes in distinct theories such that the results are still scattering amplitudes in yet a third field theory.

To illustrate these features we will focus on gauge theories with matter in adjoint representation of the gauge group. For other theories of vector fields as well as for gauge theories with matter in other representations, we refer the reader to reviews~\cite{Bern:2019prr,Adamo:2022dcm} and to the original literature.

Consider the color-dressed tree-level (super)amplitudes of a (supersymmetric) gauge theory, perhaps with matter in the adjoint representation.  By suitably multiplying and dividing by propagators, they can be generically written as a sum over diagrams with only three-point vertices:
\begin{equation}
{\cal A}^\tree_n(1,2,3,\ldots,n)\,= g^{n-2} \sum_{{\rm diags.}\, i}
                \frac{n_i c_i }{\prod_{\alpha_i} p^2_{\alpha_i}}\,.
\label{Anrep}
\end{equation}
The $c_i$ are color factors obtained by attaching an structure constant $f^{abc}$ to each vertex of the diagram\footnote{It is useful to normalize the structure constants with an extra $i\sqrt{2}$ compared to the usual textbook definitions to stay consistent with the conventions used in the amplitudes community.} and $\delta^{ab}$ to each internal line, $n_i$ are kinematic numerators which include 
the Grassmann delta function enforcing the conservation of the supermomentum and possibly further dependence on Grassmann variables, and the $p^2_{\alpha_i}$ in the denominator are the inverse propagators for the various internal lines $\alpha_i$ of the diagram $i$.
There are three $(2n -5)!!$ such diagrams with $n$ external lines: three at four-point amplitudes, fifteen for five-point amplitudes, etc.

\begin{figure}[tb]
\begin{center}
\includegraphics[width=4.in]{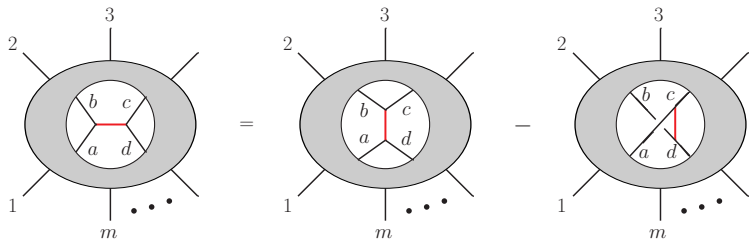}
\end{center}
\vskip -.3 cm
\caption{A Jacobi identity embedded in a generic diagram, either
at tree or loop level. 
Other than the exposed four-point tree subdiagrams, the remaining parts
of each of the diagrams is identical.  
}
\label{GeneralJacobiFigure}
\end{figure}

The two fundamental properties of structure constants is their complete antisymmetry and the Jacobi relation, 
\begin{align}
 f^{abe}f^{cde}+f^{dae}f^{cbe}+f^{bde}f^{cae}=0 \,,
 \label{jacobiGeneral}
\end{align}
i.e. the commutation relations of the gauge group generators in the adjoint representation. Since the color factors are products of structure constants, the Jacobi relation transfers to one such relation for every internal edge of the diagram. For example, the color factors of three diagrams in Fig.~\ref{GeneralJacobiFigure} are
\begin{align}
 c_i \equiv {\cdots} f^{a b e} f^{e c d}{\cdots} \,, \hskip .4 cm
 c_j \equiv {\cdots} f^{d a e} f^{e b c}  {\cdots} \,, \hskip .4 cm
 c_k \equiv {\cdots} f^{a c e} f^{e d b}  {\cdots} \,,
\end{align}
where the ellipsis stands for factors common to all three diagrams. Then, Eq.~\eqref{jacobiGeneral} becomes\footnote{In any Jacobi relation, the relative signs depend on the chosen relation between labeling of edges and structure constants. Here we chose a clockwise relation.}
\begin{equation}
 c_i - c_j + c_k=0\, ,
\label{jacobic}
\end{equation}
or, equivalently, the equality in Fig.~\ref{GeneralJacobiFigure}.

As mentioned, the duality between color and kinematics is the statement that kinematic numerators $n_i$ obey algebraic relations in one-to-one correspondence with those obeyed by the color factors $c_i$.
The antisymmetry of the structure constants implies that, under permutations of certain internal legs (e.g. legs $a$ and $b$ in the first diagram of Fig.~\ref{GeneralJacobiFigure}), The color factors are antisymmetric. Thus, imposing 
this aspect of color-kinematics duality requires that
\begin{equation}
c_i \rightarrow -c_i \hskip 1 cm  \Rightarrow
\hskip 1 cm
 n_i \rightarrow -n_i\,.
\label{SelfAntisym}
\end{equation}
Imposing the second aspect of color-kinematics duality, that kinematic numerators obey 
the dual (kinematic) Jacobi relations, is
\begin{equation}
c_i =  c_j - c_k \,  \hskip 1cm \Rightarrow
\hskip 1 cm
n_i =  n_j - n_k\, .
\label{jacobin}
\end{equation}
As emphasized in \fig{GeneralJacobiFigure}, the kinematic Jacobi identity require that lines that are common to the three participating graphs are labeled identically. 

Apart from exposing structure in the kinematic numerators of color-dressed amplitudes, the color-kinematics duality provides a means to obtain, through double copy, gravitational (super)amplitudes from gauge-theory (super)amplitudes.
As we will discuss shortly, the duality between color and kinematics relates the gauge-theory gauge symmetry to the diffeomorphism invariance of gravitational amplitudes. For this reason we may relax the definition of the duality to the requirement that kinematic numerators $n_i$ obey algebraic relations in one-to-one correspondence with those obeyed by the color factors $c_i$ {\em and required by gauge invariance}.
In particular, only the color factor relations that hold for a generic gauge group should be considered, while those valid  for special gauge groups or for particular representations, should not be imposed on the kinematic numerators. 

Color-kinematics duality suggests that, in analogy with the construction of color factors from the structure constants of the color algebra, the kinematic numerators may also be constructed from the structure constants of a {\em kinematic algebra}. Partial results are available either for special theories or for sectors of generic gauge theories~\cite{Monteiro:2011pc, OConnellAlgebras, Monteiro:2013rya,Fu:2016plh, Cheung:2016prv, Fu:2018hpu, Chen:2019ywi, Chen:2021chy, Cheung:2021zvb,Brandhuber:2021bsf, Cheung:2022mix,Ben-Shahar:2021doh,Ben-Shahar:2021zww}.

As formulated above, color-kinematics duality relates amplitudes' building blocks---kinematic numerators---that are not gauge-invariant. They can be changed through generalized gauge transformations, which effectively add arbitrary functions to amplitudes multiplied by combinations of color factors which vanish because of color Jacobi relations. 
This freedom to modify color-dressed amplitudes has a reflection on color-ordered amplitudes~\cite{BCJ}:
\begin{equation}
\sum_{i=2}^{m-1} p_1 \cdot (p_2+\ldots + p_i) \, A_m^\tree(2, \ldots , i, 1, i+1,  \ldots, m)=0\,.
\label{BCJrels}
\end{equation}
These relations, usually referred to as the {\em fundamental BCJ amplitude relations},
provide a gauge-invariant test for the existence of color-kinematics duality.

Perhaps the most remarkable aspect of color-kinematics duality is that  an amplitude---or, at loop level, the integrand of an amplitude---of a gravitational theory can be constructed by replacing the color factors of one gauge amplitude with the color-dual kinematic numerators of another:
\begin{equation}
\label{cTon}
c_i\rightarrow n_i \, .
\end{equation}
To be specific and focusing on tree-level amplitudes, given a tree-level $n$-point gauge-theory amplitude in Eq.~\ref{Anrep} together with a second $n$-point amplitude in another gauge theory, possibly with a different field content or gauge group,
then the supergravity amplitudes is
\begin{eqnarray}
&& {\cal M}^\tree_n(1,2,3,\ldots,n)\,=
  i \Big(\frac{\kappa}{2}\Big)^{n-2}  \sum_{{\rm diags.}\,i} \frac{n_i \n_i}{ \prod_{\alpha_i} p^2_{\alpha_i}} \, ,
\label{squaring}
\end{eqnarray}
where the sum runs over the same set of diagrams as in Eqs.~\eqref{Anrep} and the $\n_i$ are the kinematic numerators of the second gauge theory.  
Strikingly the linearized gauge invariance of the Yang--Mills amplitudes has been promoted to linearized diffeomorphism invariance of the gravitational amplitude.  Generalizations of this construction to theories with fields in representations other than adjoint have been discussed in Refs.~\cite{Chiodaroli2013upa, Johansson2014zca, Johansson:2015oia, Carrasco:2019yyn,Carrasco:2020ywq,Carrasco:2021ptp,Carrasco:2022jxn}.

In this form, the double-copy relation requires that at least one set of the kinematic numerators, say $\n_i$, is put in a form that manifests color-kinematics duality. Assuming that the $n_i$ also satisfies the duality between color and kinematics.  Generalized gauge transformations 
\begin{equation}
n_i \to n'_i=n_i +\Delta_i, \hskip .7 cm  \text{satisfying}  \hskip .7 cm 0 =\sum_i \frac{c_i \Delta_i }{ \prod_{\alpha_i} p^2_{\alpha_i}} \,,
\label{generalizedGT}
\end{equation}
leave the gauge-theory amplitude unchanged but can yield a representation  $n'_i$ that is no longer color-dual.  Note that in this case overall contribution to the total amplitude of the generalized gauge transformations $\Delta_i$ vanish as a result of the generic algebraic properties of the color-weights because of the algebraic identities they satisfy. Even so gravitational double-copy construction with at least one manifestly color-dual $\n_i$ can proceed and will be equal to double-copy construction using two color-dual representations following:
\be
\sum_i \frac{n'_i \n_i }{\prod_{\alpha_i} p^2_{\alpha_i}}  = \sum_i \frac{n_i \n_i }{\prod_{\alpha_i} p^2_{\alpha_i}} + \sum_i \frac{\Delta_i \n_i }{\prod_{\alpha_i} p^2_{\alpha_i}}\,.
\ee
The second term on the right vanishes as color-dual $\n_i$ satisfy the same algebraic identities as the $c_i$ in \eqn{generalizedGT}.  

 A similar argument leads to understanding how linearized gauge invariance is promoted to linearized diffeomorphism invariance.  Gauge invariance means that the gauge amplitude will be invariant under shifts of polarizations $\varepsilon_\mu(k) \to \varepsilon_\mu(k)+k_\mu$.  With $\delta_i = n_i|_{\varepsilon \to k}$, the amplitude is gauge invariant only if the shifts satisfy,
\begin{equation}
 \sum_i  \frac{c_i \delta_i}{ \prod_{\alpha_i} p^2_{\alpha_i} } = 0 \,.
\end{equation}
As the $n_i$ are independent of the color, this again relies only on the algebraic properties of $c_i$.   As a result, if $\n_i$ are color-dual, then clearly:
\begin{equation}
 \sum_i  \frac{\n_i \delta_i}{ \prod_{\alpha_i} p^2_{\alpha_i} } = 0 \,.
\end{equation}

Linearized diffeomorphisms arise from transformations of the type $\varepsilon_{\mu \nu}(k) \to \varepsilon_{\mu \nu}(k) + k_\mu q_\nu + k_\nu q_\mu$, where $q$ is a null reference momenta obeying $k_\mu q^\mu=0$. With the double-copy polarizations $\varepsilon_{\mu\nu}$ equal to the symmetric traceless part of  the product of vector polarizations $\varepsilon_\mu \varepsilon_\nu$ in the double copy, it is straightforward to see that the variation of \eqn{squaring} under linearized diffeomorphism goes as:
\be  
{\cal M}^\tree_n \to {\cal M}^\tree_n -i \left( \sum_i {\frac{ n_i |_{\varepsilon \to q} \tilde{\delta}_i }{\prod_{\alpha_i} p^2_{\alpha_j} }} +\sum_i {\frac{ \delta_i  \n_i |_{\varepsilon \to q}}{\prod_{\alpha_i} p^2_{\alpha_i} }} \right) .
\label{Mdiff}
\ee
Both terms in the second expression vanish under color-dual properties of $n_i$ and $\n_i$, leaving the double-copy amplitude invariant under linearized diffeomorphisms.

At tree level, generalized gauge transformations can be used to eliminate some of the kinematic 
numerators. The rest of the color-dual basis numerators can be expressed in terms of color-ordered partial (super)amplitudes. Color-kinematics duality, whether manifest or not, guarantees independence on the choice of kinematic numerators to be eliminated.
The net result is the Kawai, Lewellen and Tye~\cite{Kawai:1985xq} (KLT) relations, derived long ago in the context of string theory. Through five points and written here for superamplitudes, they are
\begin{align}
\calMthree(1,2,3) = \null &   i A_3^\tree(1,2,3) \widetilde A_3^\tree(1,2,3)\,, \nn \\
\calMfour(1,2,3,4) = \null &  - i  s_{12} A_4^\tree(1,2,3,4) \widetilde A_4^\tree(1,2,4,3)\,, \nn \\
\calMfive(1,2,3,4,5) =
  \null &  i\, s_{12} s_{45} A_5^\tree(1,2,3,4,5)  \widetilde A_5^\tree(1,3,5,4,2)  \nn\\
  &
   + i \, s_{14} s_{25}A_5^\tree(1,4,3,2,5) \widetilde A_5^\tree(1,3,5,2,4) \, .
\label{KLT45}
\end{align}
Starting from two gauge theories $A_n^\tree$ color-ordered partial (super)amplitudes, 
${\cal M}_n^\tree$ are tree-level amplitudes in a gravitational theory.
Here $s_{ij} = (p_i+p_j)^2$ are two-particle Mandelstam invariants. 
Explicit expressions for the KLT relations at $n$ points may be found in Ref.~\cite{MultiLegOneLoopGravity} and can be straightforwardly converted to superspace expressions.

It is, of course, important to understand what theories arise from double-copy constructions. Given a particular double-copy construction given in terms of kinematic weights of two progenitor gauge theories, the first step towards identifying what theory generates the double-copy amplitudes is to understand the speactrum and the number of manifest supercharges. The latter  follows from the observation that the kinematic numerators of each (gauge) theory contain the Grassmann delta function enforcing the conservation of the corresponding supermomentum. Each term in the double-copy amplitude~\eqref{squaring} therefore contains the product of the Grassmann delta functions,
\be
\delta^{\text{double copy}}\left(\sum \eta \lambda\right) = 
\delta^{\text{L}}\left(\sum \eta_\text{R} \lambda\right)
\delta^{\text{R}}\left(\sum \eta_\text{R} \lambda\right) .
\ee
The number of supercharges is therefore the sum of the number of supercharges of the two (gauge) theories. For example, by taking the double copying of two theories with 16 supercharges (${\cal N}=4$ supersymmetry in four dimensions) we obtain a theory with 32 supercharges (${\cal N}=8$ supersymmetry in four dimensions)

The spectrum of physical states of the double-copy theory is the tensor product of spectra of the two participating gauge theories. To illustrate this, we consider the case that these two theories are both ${\cal N}=4$ SYM theory, whose physical states are organized in the $\NeqFour$ on-shell superfield in Eq.~\eqref{Neq4multiplet}.  Distinguishing the two theories by decorating them with $L$ and $R$ subscripts,  the double-copy states are collected in Table~\ref{table_N8States}. They can be identified as the states of ${\cal N}=8$ supergravity. 
This is consistent with the previous observation that the double-copy of two theories with ${\cal N}=4$ supersymmetry yields a theory with ${\cal N}=8$ supersymmetry. 

\def\tI{{\tilde I}}
\def\tJ{{\tilde J}}
\def\tK{{\tilde K}}
\def\tL{{\tilde L}}
\begin{table}[t]
\begin{center}
\begin{tabular}{c||c|c|c|c|c}
& $\raisebox{-.3 cm}{ \vphantom{|}}  g_R{}^+$ & $f_R{}_\tI^+$ & $\phi_R{}_{\tI\tJ}$ & $f_R{}_{\tI\tJ\tK}^-$ & $g_R{}_{\tI\tJ\tK\tL}^-$     \cr
         \hline
$\raisebox{.3 cm}{ \vphantom{|}} g_L{}^+$     &   $ h^{+}$  &      $\psi^{+}_{\, \tI}$      & $A^+_{\,\tI\tJ}$  &   $\chi^{+}_{ \, \tI\tJ\tK}$   &  $\phi_{ \, \tI\tJ\tK\tL}$       \cr
$\raisebox{.3 cm}{ \vphantom{|}} f_L{}_{I}^+$  & $\psi^{+}_{I}$  &      $A^{+}_{I\, \tI}$  & $\chi^+_{I\, \tI\tJ} $ &  $\phi_{I\, \tI\tJ\tK}$    &  $\chi^{-}_{I \, \tI\tJ\tK\tL}$    \cr
$\raisebox{.3 cm}{ \vphantom{|}} \phi_L{}_{IJ}$     &   $A^{+}_{IJ}$ &  $\chi^+_{IJ\, \tI}$   & $\phi_{IJ\, \tI\tJ}$  &  $\chi^{-}_{IJ\, \tI\tJ\tK}$ &  $A^{-}_{IJ \, \tI\tJ\tK\tL }$     \cr
$\raisebox{.3 cm}{ \vphantom{|}} f_L{}^-_{IJKL}$   & $\chi^+_{IJKL}$ &      $\phi_{IJKL\, \tI}$      &   $\chi^-_{IJKL\, \tI\tJ }$ &   $A^{-}_{IJKL \, \tI\tJ\tK}$ &  $\psi^-_{IJKL \, \tI\tJ\tK\tL}$ \cr
$\raisebox{.3 cm}{ \vphantom{|}} g_L{}_{IJKL}^-$       &    $\phi_{IJKL}$  &  $\chi^{-}_{IJKL\, \tI}$   & $A^-_{IJKL\,\tI\tJ}$  &   $\psi^{-}_{IJKL \, \tI\tJ\tK}$        &  $h^{-}_{IJKL \, \tI\tJ\tK\tL}$
\end{tabular}
\vskip .3 cm 
\caption{The states of $\NeqEight$ supergravity organized via the double copy. The $SU(8)$ representations are decomposed in representations of the $SU(4)\times SU(4)$ subgroup which is manifest in the construction. 
The double-copy states can be reorganized into the standard $\NeqEight$ multiplet briefly discussed in Sect.~\ref{SuperTreeSubsection}, which has manifest $SU(8)$ symmetry and contains 256 physical states. 
}
\label{table_N8States} 
\end{center}
\end{table}

The number of supercharges uniquely defines the supergravity theory 
for ${\cal N}\ge 5$ in four dimensions. Together with the spectrum it continues 
to do so for ${\cal N}=4$ supergravity theories. 
Further Lagrangian-based information on three- and perhaps higher-point interactions 
is necessary to specify the supergravity theory for theories with fewer than 16 supercharges. For example, the ${\cal N}=2$ Maxwell-Einstein theories with a five-dimensional origin are uniquely determined by their spectrum {\em and} their three-point interactions~\cite{Gunaydin:1983bi}. Alternatively, the double-copy amplitudes can be used to construct the Lagrangian of the double-copy theory. (See also Section \ref{WebSection}).

Having understood the basics of color-kinematics duality and of the double-copy construction, we conclude the tree-level section with a discussion of how double-copy promotes the global supersymmetry of the gauge theories participating in the double copy to local supersymmetry in the double-copy theory .

\vskip -.7 cm $\null$
\subsection{Color-kinematics duality and supersymmetry}
\label{sec:ckdAndSusy}
\vskip -.2 cm 

In the case of adjoint fermions in arbitrary dimensions, the duality between color and kinematics can be shown to be equivalent to the existence of supersymmetry, as shown in Ref.~\cite{Chiodaroli2013upa}. (See also Ref.~\cite{Weinzierl:2014ava}.)  The simplest example is  Yang--Mills theory  minimally coupled to a single adjoint Majorana fermion in $D$ dimensions, described by the Lagrangian
\begin{equation}
{\cal L}={\rm Tr} \Big[-\frac{1}{4}F_{\mu\nu}F^{\mu\nu} + 
   \frac{i}{2}{\bar\psi} \s D  \psi\Big] \,.
\label{Lfermion}
\end{equation}
Of course three-point amplitudes are color-dual, and both four-gluon and two-gluon-two-fermion  amplitudes automatically respects the  duality between color and kinematics. On the other hand, as we will see, requiring the existence of a color-dual form for four-fermion amplitudes leads to a constraint on the space-time dimension.  

Consider the color-dressed amplitude,
\be
\mathbb{A}^\text{tree}_4(1 \psi,2 \psi,3 \psi,4 \psi)
 = i \left( \frac{n_s c_s}{s} +\frac{n_t c_t}{t} +\frac{ n_u c_u}{u} \right) ,
\ee
where $n_i$ are kinematic weights given in terms of spinors $\bar{u}_i$ and $v_i$ , and the color weights follow the standard Mandelstam channel graphs obeying $c_s+c_t+c_u=0$.  Explicit expressions for the $n_i$ are
\begin{align}
2\, n_s&=(\bar{u}_1\gamma_{\mu}v_2)(\bar{u}_3\gamma^{\mu}v_4)\,, \hskip 1 cm 
2 \,n_t= (\bar{u}_2\gamma_{\mu}v_3)(\bar{u}_1\gamma^{\mu}v_4)\,, \nn \\ 
2\, n_u&=(\bar{u}_3\gamma_{\mu}v_1)(\bar{u}_2\gamma^{\mu}v_4)\,.
\end{align}
The spinors satisfy  $\bar{u}_i \gamma^\mu v_j =  \bar{u}_j \gamma^\mu v_i$ and $\bar{u}_i = v_i^T {\cal C}$ for charge conjugation matrix $C$.
In dimensions where a Weyl representation can be chosen one of the terms above vanishes.

This amplitude is color-dual only if the Dirac matrices are such that
\begin{equation}
\label{ck10}
n_s+n_t + n_u=0\,.
\end{equation}
This can only be satisfied in specific dimensions, $D\in\{3,4,6,10\}$.  Interestingly, the color-dual requirement in \eqn{ck10} is exactly the supersymmetric Fierz identy necessary for \eqn{Lfermion} to be invariant under supersymmetry transformations. Similar analysis holds for pseudo-Majorana fermions.

Should this relationship between color-dual theories of adjoint fermions and supersymmetry be surprising? Perhaps not from the perspective of double-copy construction of gravitational theories.
After all, we can always double-copy adjoint spin-1/2 fermions with vectors to get spin-3/2 fermions, but such amplitudes are only consistent for a theory that can support spin 3/2 particles as part of an associated local supersymmetry multiplet.   As we have seen, the maximal dimension for a color-dual gauge theory
with adjoint fermions is  ten-dimensions, therefore bounding the highest dimension for the direct double-copy construction of maximal supergravity as the double-copy between two maximally symmetric gauge theories.  

Spin-3/2 particles belonging to a local supermultiplet, or gravitini, are an excellent case-study for how constraints of global supersymmetry lead through double-copy to constraints of local supersymmetry.
The linearized local supersymmetry transformation of a gravitino polarization vector-spinor $u{}_{\mu}^\alpha(k)$, analogous to the linearized diffeomorphism transformation of a graviton's polarization, is given,
\begin{equation}
u{}_{\mu}^\alpha(k)  \rightarrow u{}_{\mu}^\alpha(k) + k_{\mu} \xi^{\alpha} \,,
\end{equation}
where $\xi_{\alpha}$ satisfies the massless Dirac equation $k\llap/ \xi = 0$. This allows $\xi$ to preserve the $\gamma$-tracelessness of the gravitino physical states. Consider how this vector-spinor can arise through double-copy---one copy, say $\tilde{n}_i$, contributes a spinor $\tilde{u}^\alpha(k)$ and the other, $n_i$ contributes the polarization vector $\varepsilon_i(k)$ associated to a vector field. 
%
%
If color-dual numerators $\n_i$ are dependent on a spinor $\tilde{u}$, then they will remain color-dual if $\tilde{u}\to\xi$ as $\xi$ satisfies all the same properties as $\tilde{u}$. As such the supersymmetry transformation of a double-copy representation, cf.~\eqn{Mdiff},  given by,
\vskip -.2 cm 
\be
{\cal M}_m^\tree\rightarrow {\cal M}_n^\tree + \sum_i \frac{\delta_i {\tilde n}_i\big|_{{\tilde u}\rightarrow \xi} }{{\prod_{\alpha_i} p^2_{\alpha_i}}} \,,
\ee
is seen to leave the amplitude invariant as the second term vanishes.   The amplitudes of the double-copy theory respect linearized local supersymmetry.

\section{Loop-Level Methods}
\label{LoopLevelSection}

In this section we give an overview of methods that construct loop-level amplitudes from tree-level ones, focusing on those methods that have been used to compute ultraviolet counterterms at higher loops. 
The primary tools are the unitarity method~\cite{UnitarityMethod, Fusing, BernMorgan, FiveLoop} and the double copy~\cite{Kawai:1985xq, BCJ, BCJLoop, Bern:2019prr}; a key feature of these methods is that it builds into loop-level amplitudes the simplifications and structures found at tree level.  In particular, (super)gravity tree amplitudes constructed via the double copy can be immediately applied to construct loop-level amplitudes, which in turn determine the ultraviolet counterterms.
The basic strategy is the same whether we use components or on-shell superspaces~\cite{Nair,FreedmanUnitarity, Bern:2009xq, Dennen:2009vk, Bern:2010qa}.  (See Refs.~\cite{BernHuangReview, JJHenrikReview, Bern:2019prr, Bern:2022wqg} for reviews and further details.)

\vskip -1. cm $\null$
\subsection{Overview of the unitarity method}
\label{unitarityOverview}
\vskip -.2 cm 


\begin{figure}[tb]
\begin{center}
\includegraphics[width=1.9 truein]{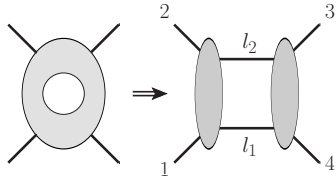}
\end{center}
\hskip -.4 cm 
\caption[a]{\small The $s$-channel two-particle cut of the one-loop four-point amplitude. Each blob represents a tree-level amplitude and each exposed line is taken to be on-shell.
}
\label{TwoParticleCutFigure}
\vskip -.6 cm
\end{figure}

Unitarity of the $S$-matrix implies that loop and tree amplitudes are interconnected via discontinuities across branch cuts. Such discontinuities, usually referred to as ``Cutkowsky cuts", are defined
as the phase-space integral over products of lower-loop (and perhaps also higher-point) amplitudes.
Provided that additional information about the high-energy behavior of amplitudes is available, these discontinuities can be used to reconstruct amplitudes via dispersion integrals.

The modern unitarity method builds on these ideas and on the additional information that (in principle) loop amplitudes can be given in terms of Feynman diagrams, that is that they are integrals of rational functions with well understood properties, referred to as ``integrands". 
The structure of Feynman diagrams implies that the residues of these rational functions are given in terms of unintegrated products of  tree-level amplitudes or of the rational functions corresponding to lower-loop (perhaps higher-point) amplitudes summed over the possible physical states. We refer to these unintegrated products of trees as ``generalized cuts" or simply ``cuts".
For any amplitude these generalized cuts are sufficient to reconstruct its integrand.

To illustrate the basics of this method let us discuss the construction of a one-loop amplitude with only massless particles. Such an amplitudes is fully constructible from the two-particle (generalized) cuts. One of them is shown in \fig{TwoParticleCutFigure} for the $s=(p_1+p_2)^2$ cut and additional inequivalent ones may be obtained by relabeling the external particles.  
The generalized $s$-channel cut in \fig{TwoParticleCutFigure} is\footnote{This discussion holds for any theory with four-point tree amplitudes. We may, equally well choose ${\cal N}$-extended SYM theory, or the NLSM, or some supergravity theory, etc.}
\begin{equation}
C_s = \sum_{\rm states}  A^\tree(-l_1, 1,2, l_3) \, A^\tree(-l_3, 3,4, l_1) \,,
\label{SCut}
\end{equation}
where the sum runs over all physical states in the theory. 
The momenta of all exposed lines, in particular those of internal lines are placed on shell, $l_i^2 = 0$, which corresponds to evaluating the residue of the singularity corresponding to those propagators being on shell. 
The input amplitudes can be either in component form (and the state sum is carried out explicitly) or in superspace form~\cite{Nair,FreedmanUnitarity, Bern:2009xq, Dennen:2009vk, Bern:2010qa} where the state sum is carried out through Grassmann integration, in a similar fashion to \eqn{Supersum} at tree level.

Within the context of the unitarity method, there are various strategies for building multi-loop amplitudes from tree-level amplitudes.  In particular,  one may consider cuts recursively where only a single internal line is placed on shell~\cite{NigelGlover:2008ur, Bierenbaum:2010cy, Caron-Huot:2010fvq}, which have been used to construct multiloop planar integrands of $\NeqFour$ SYM theory~\cite{Arkani-Hamed:2010zjl}.  
Another strategy, known as prescriptive unitarity~\cite{Bourjaily:2017wjl}, has been used to systematically build two-loop amplitudes with an arbitrary number of external legs.  

\begin{figure}[tb]
\begin{center}
\includegraphics[width=4.6 truein]{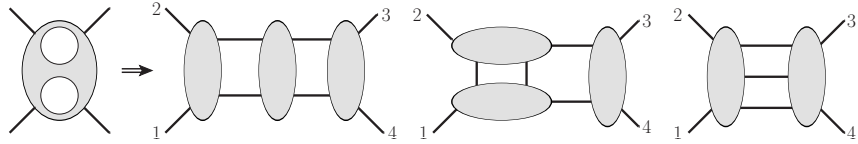}
\caption[a]{\small A sample spanning set of generalized cuts for a massless two-loop four-point amplitude.  The complete set is given by the independent relabelings of the four external legs.  Each blob represents a tree amplitude and the exposed intermediate lines are on shell. 
  }
\label{TwoLoopFourPtFigure}
\end{center}
\end{figure}

To construct a complete multiloop integrand one needs a ``spanning set'' of cuts.  Such sets have the property that any potential independent contribution is determined by at least one of the generalized cuts. By finding an integrand whose generalized cuts match the appropriate spanning set, one is guaranteed to find the complete integrand. 
The simplest example of such a spanning set is that for the one-loop four-point massless amplitude, shown in \fig{TwoParticleCutFigure}. Any part that is not in this cut (or its relabelings) has a single propagator and thus vanishes in dimensional regularization.
A less trivial example is the spanning set of cuts for a massless two-loop four-point amplitude.  It is illustrated in \fig{TwoLoopFourPtFigure}, where the complete set is obtained by including all independent relabelings of external legs.  

\begin{figure}[tb]
\begin{center}
\includegraphics[width=4 truein]{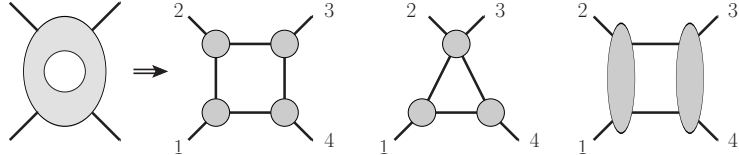}
\caption[a]{\small A sample spanning set of generalized cuts for a massless one-loop four-point amplitude in the method of maximal cuts. The complete set is given by the independent relabelings of the four external legs.}
\label{OneLoopFourPtMaxCutsFigure}
\end{center}
\end{figure}

Here we focus on the method of maximal cuts~\cite{FiveLoop}, which has been central to computing ultraviolet divergences in supergravity theories, described in the next section. 
This method uses an overcomplete spanning set of generalized cuts.
One begins with the cuts where the maximum number of propagators are placed on shell~\cite{FiveLoop} (the ``maximal cuts") and systematically considers all generalized cuts with increasingly fewer cut lines, thus building an integrand that matches all generalized cuts.
The resulting integrand is naturally organized in terms of graphs with increasingly higher-multiplicity vertices, referred to as ``contact terms". 
The process terminates when the only remaining potential contact terms exceed power counting constraints of the theory or integrate to zero in dimensional regularization.
From this perspective, the building blocks for loop amplitudes are sums of products over $m$ tree amplitudes,
\begin{equation}
C \equiv \sum_{\rm states} A^\tree_{(1)} A^\tree_{(2)} A^\tree_{(3)} \cdots A^\tree_{(m)} \,,
\end{equation}
where the sum runs over all physical states that can cross the cuts.  As for \eqn{SCut} the state sum can be carried out either in components or in an on-shell superspace. The amplitudes can either be either gauge or gravity amplitudes.  Moreover, in gauge theories they can be either color ordered or color dressed. 
As a simple example, at one loop the overcomplete spanning set of cuts used in the maximal cut method is shown in \fig{OneLoopFourPtMaxCutsFigure}. 

For the case of maximal supersymmetry there exists a large number of variety of additional methods and shortcuts for obtaining contributions efficiently, and is related to the relatively simple structure of the amplitudes~\cite{FiveLoop, Cachazo:2008dx, Cachazo:2008vp, Bern:2010tq, Arkani-Hamed:2010zjl, Arkani-Hamed:2010pyv, Carrasco:2021otn}.  This makes it possible to carry out explicit calculations, even at five loops.

A convenient approach in practical calculations, especially at high loop orders, it is to first constructing an ansatz for the diagram numerators containing all possible independent momentum-dependent terms, each with an arbitrary coefficient to be determined by matching the generalized cuts.  
When using superamplitudes,  the numerators will also depend on Grassmann parameters. To simplify the ansatz it is also convenient to enforce certain auxiliary constraints, such as  manifest crossing symmetry, an upper bound on the number of loop momentum factors in each term and manifest color-kinematics duality.  
If the ansatz fails to match the generalized cuts, then the ansatz needs to be enlarged by removing auxiliary constraints.
This procedure is most straightforward for the four-point amplitudes of $\NeqFour$ SYM theory, where it turns out that the ratio between the loop integrand and the tree amplitudes is a rational function of Lorentz-invariant scalar products, at least through five loops and likely beyond this~\cite{BRY,Bern:2006ew, FiveLoop,Bern:2010tq,FiveLoopN4}.  Higher-point loop  amplitudes are no longer proportional to the corresponding tree amplitudes~\cite{UnitarityMethod, Fusing, FivePointN4BCJ}; because of this, the numerator ansatz must also contain Lorentz-invariant products of momenta and polarization vectors or tensors, in addition to products of momenta.

The generalized unitarity method covers equally well both gauge and gravitational theories. However, it is generally more efficient to construct loop amplitudes in gravitational theories by first constructing corresponding gauge-theory loop integrands that satisfy the duality between color and kinematics and then using the double copy~\cite{BCJLoop}.   
At five-loop order, where it is difficult to find such duality-satisfying representations, the generalized double copy provides a means to judiciously use the double copy to promote gauge-theory loop integrands to gravitational ones. We will review this in Sect.~\ref{GeneralizedDoubleCopySection}.

\vskip -.7 cm $\null$
\subsection{Example: The one-loop four-point integrand of $\NeqFour$ SYM theory}
\vskip -.2 cm 

To illustrate the generalized unitarity method we discuss a simple example: the on-shell superspace form of the one-loop four-point integrand of  $\NeqFour$ SYM theory, originally discussed in Refs.~\cite{Drummond:2008bq, ArkaniHamed2008gz, FreedmanUnitarity, Bern:2009xq}. 
As discussed, the spanning set of generalized supercuts contains a single element, $C_s$, and its permutations, as shown in  \fig{TwoParticleCutFigure}. We will interpret it as a color-ordered supercut, so the superamplitudes used to construct it are color-ordered superamplitudes.
The sum over states is obtained by carrying out the Grassmann integrals,
\begin{equation}
C_s=  \int d^4\eta_{l_1}\int d^4\eta_{l_3}  \,
{A}_4^{\rm MHV}(-l_1,1,2, l_3) \,
{A}_4^{\rm MHV}(-l_3,3,4, l_1)\,,
\label{OneLoopCutSuperExample}
\end{equation}
where the MHV superamplitudes are given by relabeling \eqn{MHVsYM} for $n=4$.  
In simple cases it is straightforward to evaluate the state sums by direct Grassmann integration, but for more complex cuts it is useful to either use index diagrams that track all internal-state configurations, or to systematically solve the system of linear equations implied by the Grassmann delta functions~\cite{Bern:2009xq}.  Here we illustrate the latter method and refer the reader to Ref.~\cite{Bern:2009xq} for further details on the definition and use of index diagrams.

Regularization, usually in the form of dimensional regularization, is needed to properly recover the IR and ultraviolet properties of the theory. To this end, the construction of the integrand must be done in an arbitrary dimension $D$. It is often convenient however to first construct the integrand in $D=4$ dimensions and subsequently verify its validity in an arbitrary dimension and suitably modify it if necessary. 
We will return to this point in Sect.~\ref{sec:regularization}; in this section we will concern ourselves with the four-dimensional integrand of the four-point one-loop amplitude of ${\cal N}=4$ SYM theory, which turns out to be the same for all $D<10$.

The $\eta$ integration acts on the two supermomentum delta functions contained in the tree-level superamplitudes entering the cut \eqref{OneLoopCutSuperExample}, 
\begin{equation}
\delta^{(8)}\Bigl(\lambda_{l_3}^\alpha\eta_{l_3}^a -\lambda_{l_1}^\alpha\eta_{l_1}^a
             +\lambda_{1}^\alpha\eta_{1}^a+\lambda_{2}^\alpha\eta_{2}^a \Bigr) \,
\delta^{(8)}\Bigl( \lambda_{l_1}^\alpha\eta_{l_1}^a -\lambda_{l_3}^\alpha\eta_{l_3}^a
             +\lambda_{3}^\alpha\eta_{3}^a+\lambda_{4}^\alpha\eta_{4}^a \Bigr) \,.
\end{equation}
We may interpret the delta functions as imposing a set of linear equations on the $\eta_i^a$ that we solve. Some of these constraints represent the overall supermomentum conservation, and can always be isolated by taking suitable linear combinations of the arguments of the delta functions.
In the case at hand, this is realized by simply adding the argument of the first delta function to the argument of the second one.
Since this delta function does not depend on a the integration variable it can be taken outside of the Grassmann integral.  The cut \eqref{OneLoopCutSuperExample} then becomes
\begin{equation}
C_s=-\delta^{(8)}\Bigl(\sum_{i=1}^{4}\lambda_{i}^\alpha\eta_{i}^a\Bigr)
\frac{ \int d^4\eta_{l_1}d^4\eta_{l_3}\, \delta^{(8)}\Bigl(\lambda_{l_3}^\alpha\eta_{l_3}^a
             -\lambda_{l_1}^\alpha\eta_{l_1}^a
             +\lambda_{1}^\alpha\eta_{1}^a+\lambda_{2}^\alpha\eta_{2}^a \Bigr)}
{ \spa{l_1}.1 \spa1.2\spa2.{l_3}\spa{l_3}.{l_1} \, \spa{l_3}.3 \spa3.4 \spa4.{l_1} \spa{l_1}.{l_3}}\,. 
\label{FourPtGrassmannInt}
\end{equation}
The Grassmann integrals can now be evaluated one by one for each  R-symmetry index.  Choosing for example $a=1$, the fermionic integration is,
\begin{equation}
\int d\eta_{l_1}^a \, d\eta_{l_3}^a\delta^{(2)}
         \Bigl(\lambda_{l_3}^\alpha\eta_{l_3}^a
             -\lambda_{l_1}^\alpha\eta_{l_1}^a
             +\lambda_{1}^\alpha\eta_{1}^a
             +\lambda_{2}^\alpha\eta_{2}^a \Bigr)= -\langle l_3 \, l_1\rangle\,,
\end{equation}
which follows from integrating the form of delta function in \eqn{DeltaSpinor}.
The other three cases $a=2,3,4$, similarly, give the same factor.
Thus, the Grassmann integration gives a factor of $\langle l_3\, l_1\rangle^4$.

In more complicated cases it is more efficient to approach the Grassmann integration in \eqn{FourPtGrassmannInt}  more systematically, viewing the problem as solving a system of linear constraints, 
 \begin{equation}
\lambda_{l_1}^\alpha\eta_{l_1}^a-\lambda_{l_3}^\alpha\eta_{l_3}^a
 =\lambda_{1}^\alpha\eta_{1}^a+\lambda_{2}^\alpha\eta_{2}^a \,, \hskip 1 cm  a=1,\dots ,4 \,. 
\end{equation}
There are a total of eight constraint matching the eight integration
variables, $\eta_{l_1}^a$ and $\eta_{l_3}^a$, which are therefore completely fixed.  
The integral is then given by the Jacobian of the matrix of the coefficients of the linear 
equations,
\begin{equation}
J =  \det{}^{\! 4} \left|
\begin{array}{cc}
    \lambda_{l_1}^1 &\; - \lambda_{l_3}^1 \\[8pt]
    \lambda_{l_1}^2 & \; - \lambda_{l_3}^2 
\end{array}
\right| = \spa{l_1}.{l_3}^4 ,
\label{GrassmannResult}
\end{equation}
which matches the result obtained above by evaluating the Grassmann integrals one by one.

Thus, the $s$-channel cut \eqref{OneLoopCutSuperExample} of the four-point superamplitude is
\begin{equation}
{\cal C}_s = 
- \delta^{(8)}\Bigl(\sum_{i=1}^n\lambda_i^\alpha\eta_i^a\Bigr) \,\frac{\spa{l_1}.{l_3}^4}
{ \spa{l_1}.1 \spa1.2\spa2.{l_3}\spa{l_3}.{l_1}
\, \spa{l_3}.3 \spa3.4 \spa4.{l_1} \spa{l_1}.{l_3}} \,. \hskip .5 cm
\end{equation}
To put this into a form reminiscent of the result obtained from Feynman diagrams,
we rationalize the denominators using, for example,
\begin{equation}
 \frac{1}{\spa2.{l_3}} = - \frac{\spb2.{l_3}}{(p - k_1)^2} \, ,
\label{Rationalize}
\end{equation}
where we set $l_1 = p$, $l_3 = p - k_1 - k_2$ and we used the on-shell
conditions $l_1^2=l_3^2=0$.  These simplifications lead to
\begin{equation}
C_s =  i {A}_4^{\rm MHV}  \frac{ N_s}{(p - k_1)^4 (p + k_4)^4}  \,,
\label{SCutSusyB}
\end{equation}
where the numerator $\cal N$ is given by
\begin{equation}
N_s  = \spb{l_1}.1 \spa1.4 \spb4.{l_1} \spa{l_1}.{l_3}
             \spb{l_3}.3 \spa3.2 \spb2.{l_3} \spa{l_3}.{l_1} \,.
\end{equation}
The evaluation of such products of spinor products proceeds by using the spinor completeness relation~\cite{ManganoParkeReview} 
$| p  \rangle\, [ p| = {\textstyle \frac{1}{2}} (1 + \gamma_5) \s p$ and
$| p ]\, \langle p| =  {\textstyle \frac{1}{2}} (1 - \gamma_5) \s p$ ,
to turn $N_s$ into a trace of products of Dirac matrices multiplied by momenta. Evaluating 
this trace through textbook methods and further using the on-shell and cut conditions leads to the simplified expression for the cut $C_s$ in Eq.~\eqref{SCutSusyB}:
\begin{eqnarray}
C_s={A}_4^{\rm tree}(1,2,3,4)\, \frac{-ist}{(p-k_1)^2(p+k_4)^2}\,.
\label{D4CutResult}
\end{eqnarray}
Including the cut propagators we arrive at the four-point one-loop color-ordered superamplitude,
\begin{equation}
{A}_4^{\rm1-loop}(1,2,3,4)= -st {A}^{\rm tree}_4 I_4(s,t) \,,
\end{equation}
%
where $I_4(s,t)$ is the scalar box integral,
\begin{equation}
I_4(s,t) = -i \int \frac{d^{4-2\eps}p}{(2\pi)^{4-2\eps}} \, \frac{1}{p^2 (p - k_1)^2 (p - k_1 - k_2)^2 (p + k_4)^2} \,.
\label{BoxIntgegral}
\end{equation}
An important consistency condition is that the identical expression follows by analyzing the $t$-channel cut. It is quite remarkable that all integrals other than the scalar box integral cancel out, a fact that was understood long ago by considering the low-energy limit of string theory~\cite{Green:1982sw}.

In the example discussed above, the only remaining dependence on Grassmann variables is in the overall delta function enforcing the conservation of the supermomentum. This structure is special to MHV superamplitudes. 
In more general amplitudes more fermionic variables remain after performing all Grassmann integrations.
We note that the overall supermomentum delta-function is sufficient to prove the finiteness of $\mathcal N \ge 2$ SYM amplitudes in four dimensions to all loop orders.  
For supergravity the overall Grassmann delta function is insufficient to prove finiteness because the dimensionful coupling constant implies an increasingly worse behavior in the absence of hidden symmetries. 

The methods outlined and illustrated here have been used to construct amplitudes in $\NeqFour$ SYM theory and $\NeqEight$ supergravity through five loops~\cite{BRY, BDDPR,Bern:2007hh, Bern:2008pv, Bern:2009kd, Bern:2010tq, FivePointN4BCJ, SimplifyingBCJ, FiveLoopN4,GeneralizedDoubleCopyFiveLoops, UVFiveLoops, Henn:2019rgj} as well as in a variety of other theories with less than maximal supersymmetry~\cite{Bern:2013yya, Chiodaroli2014xia, Mogull:2015adi, Chiodaroli:2017ngp}.  
The double copy played a central role in these calculations, either for simplifying the gravitational tree amplitudes that enter the cuts, or for finding complete gravitational integrands as double copies of gauge-theory integrands.


\vskip -.7 cm $\null$
\subsection{Loop-level duality between color and kinematics}
\vskip -.2 cm 

In Sect.~\ref{cktrees} we discussed the duality between color and kinematics for tree-level amplitudes. Since the generalized unitarity method constructs loop integrands from tree amplitudes, it is natural to conjecture that the duality holds to all loop orders.
While no formal proof has been constructed, many examples are available.
Similar to the tree-level organization of amplitudes, any amplitude to any loop order can be written in terms of trivalent graphs; higher-point interaction vertices (or ``contact terms") commonly present in Feynman diagrams are put in this form by multiplying and 
dividing by appropriate propagators.
For any field theory with fields in the adjoint representation, $L$-loop $m$-point amplitudes can be written as
\begin{equation}
 {\cal A}^{L-\rm loop}_m =  {i^L} {g^{m-2 +2L }}
\sum_{{\cal S}_m} \sum_{j}{\int \prod_{l=1}^L \frac{d^{D} p_l}{ (2 \pi)^{D}} \frac{1}{S_j}  \frac {n_j c_j}{\prod_{\alpha_j}{p^2_{\alpha_j}}}}\, ,
\label{LoopGauge}
\end{equation}
where second sum runs over all distinct $m$-point $L$-loop graphs with only trivalent vertices, labeled by $j$,  the first sum runs over the set ${\cal S}_m$  of $m!$ permutations of external legs, and $S_j$ is the symmetry factor of graph $j$ and removes the overcount due to the automorphisms of this graph.
For each term, there are $L$ $D$-dimensional integrals over the independent loop momenta and the denominator is given by the product of all (inverse) Feynman propagators of the corresponding graph.
Similarly to tree-level amplitudes, the color factor $c_j$ is determined by the graph by associating to each vertex a gauge-group structure constant in a chosen (e.g. clockwise) labeling of lines. The kinematic numerators $n_j$ are polynomials in Lorentz-invariant products of momenta, polarization vectors and spinors and, for superamplitudes, also in Grassmann parameters.
As at tree level, this amplitude is said to obey the duality between color and kinematics if the numerator factors have the same algebraic properties as those properties of the color factors -- antisymmetry~\eqref{SelfAntisym} and Jacobi relations~\eqref{jacobin} -- 
that are required by gauge invariance.
Generalized gauge transformations~\eqref{generalizedGT} are critical for obtaining kinematic numerators with these properties~\cite{BCJ, BCJLoop, Square, JConstraintsTye}.

Although at the time of this writing the duality is a conjecture at loop level, it is supported by a number of nontrivial examples. In particular, in $\NeqFour$ SYM theory, superamplitudes manifestly obeying the duality have been 
constructed at at four points through four loops~\cite{BCJLoop,SimplifyingBCJ},
at five points at one and two loops~\cite{FivePointN4BCJ, Johansson:2017bfl} and at six and seven points at one loop~\cite{Bjerrum-Bohr:2013iza, He:2015wgf, He:2017spx, Edison:2022jln}.
In ${\cal N}=1$ and nonsupersymmetric Yang--Mills theory, some four-point amplitudes obeying the duality at one and two loops may be found in e.g.~Refs.~\cite{BCJLoop, Carrasco:2012ca, Chiodaroli2013upa, Bern:2013yya, Mogull:2015adi, Porkert:2022efy}.
Similarly, color-kinematics-satisfying representations of form factors in various supersymmetric gauge theories were found e.g.~Ref.~\cite{Yang:2016ear, Lin:2021kht, Li:2022tir} and color-kinematics-satisfying representations of matrix elements of the open superstring effective field theory were discussed in e.g.~Ref.~\cite{Edison:2022smn}.

As at tree level, gauge-theory loop-level integrands manifestly obeying color-kinematics duality can be used to obtain gravitational integrands through the double-copy construction~\cite{BCJ,BCJLoop}, that is by substituting the color factors $c_i$ of an integrand with thew kinematic numerators $n_i$ of another integrand, $c_i \rightarrow \n_i$ cf. Eq.~\eqref{cTon}, at the same loop order and multiplicity. That is, at $L$-loops and for $m$ external lines, 
\begin{equation}
{\cal M}^{L-\rm loop}_m =  {i^{L+1}} {\biggl(\frac{\kappa}{2}\biggr)^{m-2+2L}} \,
\sum_{{\cal S}_m} \sum_{j} {\int \prod_{l=1}^L \frac{d^{D} p_l}{(2 \pi)^{D}}
 \frac{1}{S_j} \frac{n_j \n_j}{\prod_{\alpha_j}{p^2_{\alpha_j}}}} \, .
\hskip .7 cm
\label{DoubleCopy}
\end{equation}
Tree-level double copy guarantees that all generalized unitarity cuts of the resulting integrand reproduce the correct expression obtained by multiplying tree-level amplitudes and summing over intermediate states, cf. Sect.~\ref{unitarityOverview}. For the same reason as at tree level, the double-copy formula \eqref{DoubleCopy} holds even if only one of the two sets of numerators, $n_j$ or $\n_j$, satisfies the duality manifestly. It is however necessary that it should be possible to bring the second set of numerators to a color-kinematics-satisfying form through generalized gauge transformations.

\vskip -.7 cm $\null$
\subsection{Generalized double copy}
\label{GeneralizedDoubleCopySection}
\vskip -.2 cm 


Availability of gauge-theory integrands that satisfy color-kinematics duality manifestly guarantees a direct path the corresponding gravitational integrands. 
The former are however not always straightforward to construct. In some cases such as the five-loop four-point $\NeqEight$ supergravity amplitudes, it has proven difficult to find such a representation. In other cases, such as the all-plus two-loop five-gluon amplitude in pure-Yang--Mills theory, the BCJ form of the amplitude has a superficial power-count much worse than that of Feynman diagrams~\cite{Mogull:2015adi} and thus an analysis of ultraviolet properties of its double copy would more cumbersome than if the corresponding amplitudes were constructed directly from e.g. generalized unitarity. 

A double-copy method that directly converts generic representations of gauge-theory amplitude to gravity ones was developed in Ref.~\cite{GeneralizedDoubleCopy} and applied in Refs.~\cite{GeneralizedDoubleCopyFiveLoops, UVFiveLoops} to construct the five-loop four-point integrand of $\NeqEight$ supergravity and study its ultraviolet properties after integration. 

The starting point is local representations of the two gauge-theory amplitudes that use only cubic graphs, as in \eqn{LoopGauge}. The next step is to naively the double-copy substitution \eqref{cTon} to these amplitudes. In general, {\em does not} result in a correct gravitational amplitude;  nevertheless, this ``naive double copy'' can be systematically improved to become the correct amplitude. 
The naive double-copy amplitude reproduces the maximal and next-to-maximal cuts of the expected (super)gravity amplitude whenever these cuts have only three-point and only one four-point tree-amplitude factors, respectively. This is because, if it is in principle possible to construct color-kinematics-satisfying representation of gauge-theory amplitudes, then the three-point and four-point tree amplitudes manifestly obey the duality~\cite{BCJ}.
Beyond the next-to-maximal cuts, the naive double copy will generally {\em not} give correct unitarity cuts, and nontrivial corrections are necessary.  These improvement terms are derived by finding the generalized gauge transformations that put the relevant trees entering the unitarity cuts in a color-kinematics-satisfying form and expressing them in terms of the (cut) numerators of the original gauge-theory amplitudes.

\begin{figure}[tb]
\begin{center}
\includegraphics[width=1.3 truein]{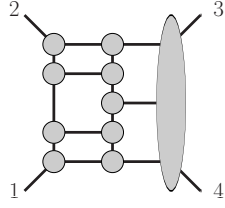}
\caption[a]{\small A sample generalized cut for the five-loop four-point decomposing the amplitude into three point amplitudes and a single five-point tree amplitude.  Each blob represents a tree-level amplitude and the exposed lines are all on shell.} 
\label{FiveLoopCutExampleFigure}
\end{center}
\vskip -.5 cm
\end{figure}

To illustrate these ideas, we discuss in some detail the generalized unitarity cut that decomposes a multiloop amplitude into sums over states of a product of three-point tree amplitudes and a single five-point amplitude, as in \fig{FiveLoopCutExampleFigure}.  
This cut can be expressed as a sum over 15 diagrams, corresponding to the 15 diagrams with only cubic vertices that make up the five-pointy tree amplitude,
\begin{equation}
{\cal C}_{5}  =\sum_{i=1}^{15} \frac{n_{i} \tn_{i}} {d^{(1)}_{i} d^{(2)}_{i} }
 + {\cal E}_{5} \,.
\end{equation} 
The first term is the naive double copy and the second is improvement term which turns out to be~\cite{GeneralizedDoubleCopyFiveLoops} 
\begin{equation}
 {\cal E}_{5}  = - \sum_{i=1}^{15}
 \frac{ \Delta_{i_1, i_2} {\tilde \Delta}_{i_1, i_2}}
  { d^{(1)}_{i} d^{(2)}_{i}} = - \frac{1}{6}\sum_{i=1}^{15}
 \frac{J_{\{i,1\}} \tJ_{\{i,2\}} + J_{\{i,2\}} \tJ_{\{i,1\}} }
  { d^{(1)}_{i} d^{(2)}_{i}} \, ,
\label{Extra5}
\end{equation}
where $\Delta$ and $\tilde\Delta$ are the generalized gauge parameters that put the cut
gauge-theory numerators in color-kinematics-satisfying form. 
The $J_{\{i,j\}}$ and  $\tJ_{\{i,j\}}$ are the referred to as ``BCJ discrepancy functions" and encode violations of the kinematic Jacobi relations and the  $1/d^{(j)}_{i}$ is the $j$th Feynman propagator of the $i$th diagram.  
Each $J_{i,j}$ consists of a sum, with appropriate signs, over the three numerator factors forming a Jacobi triplet, e.g.
\begin{equation}
J_{\{1,1\}} = n_1 + n_2 + n_3 \, .
\end{equation}
They vanish identically if the duality between color and kinematics is manifest.  

In this way we can directly connect the corrections to the naive double copy to violations of the duality. See Ref.~\cite{GeneralizedDoubleCopyFiveLoops} for further details, including a scheme for labeling and tracking the appropriate triplets.
Using such correction formulae, the naive double copy can be systematically promoted to a complete (super)gravity amplitude.  

This procedure has been used to construct the complete five-loop four-point $\NeqEight$ supergravity loop amplitude in Ref,~\cite{ GeneralizedDoubleCopyFiveLoops}.  The ultraviolet properties were described in Ref.~\cite{UVFiveLoops} and are summarized in the next section.


\vskip -.7 cm $\null$
\subsection{Comments on regularization}
\label{sec:regularization}

Scattering amplitudes in general gauge and gravitational theories are both infrared and ultraviolet divergent and thus need to be regularized. Dimensional regularization, in which amplitudes are evaluated in $D=4-2\epsilon$ dimension, is the preferred, including in massless theories~\cite{tHooft:1972tcz}, because of its flexibility and also because it preserves gauge invariance despite the care needed to preserve supersymmetry~\cite{Siegel:1979wq,FDH,FDH2}. \footnote{In planar $\NeqFour$ SYM theory only infrared divergences are present, and in this case a massive Higgs regulator has proven to be convenient because it preserves dual conformal invariance~\cite{Alday:2009zm}. }
In this scheme, divergences appear as $\epsilon^{-n}$ for positive integers $n$ and because of their presence the finite parts of amplitudes depend on terms in the integrand that are ${\cal O}(\epsilon)$, as discussed in~\cite{BernMorgan}.

Dimensionally-regularized one-loop scattering amplitudes in massless supersymmetric theories are fully determined by their generalized cuts in four dimensions~\cite{UnitarityMethod}. This is directly related to their better power counting compared to their nonsupersymmetric counterparts. 
In general, this no longer holds at higher loops, so the construction of the integrand must capture all terms to sufficiently high order in the dimensional regulator. Thus, integrands constructed from generalized cuts evaluated with four-dimensional helicity methods may need to be supplemented with further terms that vanish in four dimensions.
Explicit calculations show that four-dimensional methods give the complete integrands for four-point amplitudes in $\NeqFour$ SYM theory through six loops~\cite{BDDPR, Bern:2006ew, Bern:2007hh,FiveLoop,Bern:2010tq,Bern:2010qa,FiveLoopN4}, but not for higher-point amplitudes~\cite{Bern:2008ap}. The double copy then implies that four-dimensional methods are sufficient for the corresponding four-point amplitudes in $\NeqEight$ supergravity.

An interesting possibility is that all terms not determined by four-dimensional massless cuts have a predictable universal structure, at least near four dimensions. Until such a structure is proven and a strategy to evaluate these additional terms from four-dimensional data is devised, only $D$-dimensional calculations guarantee the completeness of the resulting amplitude.
In general, using $D$-dimensional loop momenta~\cite{BernMorgan, Bern:2000dn} makes calculations significantly more involved, because the powerful four-dimensional superspace~\cite{Nair, FreedmanUnitarity, Bern:2009xq, Dennen:2009vk, Bern:2010qa} and spinor methods~\cite{SpinorHelicityBerends, SpinorHelicityKleissStirling, SpinorHelicityXZC} can no longer be used on the intermediate states in the cuts.  This is mitigated by the six-dimensional helicity~\cite{Cheung:2009dc} and superspace~\cite{Dennen:2009vk} formalisms that we briefly reviewed in Sect.~\ref{SuperTreeSubsection}.  
They have been used to directly confirm that the four-loop four-point integrand of $\NeqFour$ SYM and the associated $\NeqEight$ supergravity integrand obtained via double copy are consistent with unitarity in dimensional regularization~\cite{Bern:2010qa}.

\section{Ultraviolet properties of supergravity Theories}
\label{UVSection}

\vskip -1.1 cm $\null$
\subsection{Symmetry constraints on counterterms}
\vskip -.2 cm

\begin{table}[t]
\begin{center}
\begin{tabular}[t]{c|c|c|c}
\; Theory & \; Counterterm \; &  \; Loop Order  \; &  \; divergence\\[2pt]
\hline \\ [-6pt]
$\; D=4, \;Q = 32,\; \NeqEight\;$ & ${\cal D}^8 R^4$ & 7 & unknown \\[2pt]
$\; D=4,\; Q = 16,\; \NeqFour\;$ & $R^4$ & 3 & no \\[2pt]
$\; D=4,\; Q = 20, \;\NeqFive\;$ & ${\cal D}^2R^4$ & 4 & no \\[2pt]
$\; D=24/5,\; Q = 32\;$ & ${\cal D}^8R^4$ & 5 & yes\\[2pt]
$\; D=5,\; Q = 16\;$ & $R^4$ & 2 & no \\[2pt]
\end{tabular}
\end{center}
\caption{Counterterms corresponding to the first potential divergence that satisfy all proven supersymmetry and
  duality-symmetry constraints~\cite{Green:2010sp, Bjornsson:2010wm, Beisert2010jx,
    Bossard:2011tq, Bossard:2013rza, Freedman:2018mrv}.
   The number of supercharges is $Q$ and $D$ is the space-time dimension.}
\label{GoodCounterTermsTable}
\end{table}
The question of which counterterms satisfy all known symmetries is still a nontrivial question despite considerable effort to answer it.  In the early 1980s the consensus opinion was that the three loop $R^4$ was valid~\cite{Green:1982sw, Marcus:1984ei, Howe:1988qz}.\footnote{By assuming the existence of an extended superfield formalism that manifests all supersymmetries one can raise this bound to seven loops~\cite{Grisaru:1982zh}; later it was shown that the counterterm cannot be written as a full superspace integral~\cite{Bossard:2011tq}.} The restrictions supersymmetry and duality symmetry impose on counterterms have been studied in great detail over the years. 

The three methods that have been applied to constrain the counterterms that can appear are:
\begin{enumerate}
    \item Use extended off-shell superpaces together with duality symmetries~\cite{Bossard:2010bd,Green:2010sp, Beisert2010jx, Green:2010sp,  Bossard:2011tq, Freedman:2018mrv}.
    \item Use Berkovits pure spinor formalism~\cite{Berkovits:2000fe} to expose the full supersymmetry~\cite{Bjornsson:2010wm}. 
    \item Use maximal cuts~\cite{FiveLoop} of amplitudes to expose the minimum powers of loop momenta required in any covariant Feynman-like diagrams~\cite{Bern:2014sna, Bern:2017lpv}.  This predicts where divergences appear, assuming no further cancellations between diagrams.
\end{enumerate}
These three methods lead to the same constraints on counterterms for the various theories displayed in \tab{GoodCounterTermsTable}. The last of these methods implies that cancellations that occur beyond this 
cannot be manifested diagram by diagram.  By definition, any further cancellations beyond the ones exposed by the above methods are {\it enhanced ultraviolet cancellations}~\cite{Bern:2014sna}.  \tab{GoodCounterTermsTable} displays the first available counterterms that satisfy all known symmetry constraints for various theories of interest. In the table, $D$ is the space-time dimensions and $Q$ is the number of supercharges.

For $\NeqEight$ supergravity in $D=4$, as demonstrated in various studies, a 
counterterm is allowed at $L=7$ loops~\cite{Green:2010sp, Bjornsson:2010wm, Bossard:2010bd, Beisert2010jx, Green:2010sp}.  By reducing the number of supercharges the loop order where the first viable counterterm exists is lowered.  In particular, 
in $D=4$ $\NeqFour$ supergravity the first counterterm is allowed at $L=3$ loops.
For $\NeqFive$ supergravity the loop order is raised to $L=4$ loops~\cite{Bossard:2011tq, Freedman:2018mrv}. As explained in Ref.~\cite{Bossard:2011tq}, these counterterms  cannot be written as full-superspace integrals, but they do appear to respect all known standard-symmetry considerations.  By increasing the space-time dimensions, one can also lower the loop order at which a divergence can first appear.  For example, half-maximal 16-supercharge supergravity in $D=5$ exhibits a possible two-loop counterterm invariant under all known symmetries~\cite{Bossard:2013rza, Bossard:2013rza}.   
By taking an unphysical dimension of $D=24/5$ for maximal 32-supercharge supergravity, corresponding to $\NeqEight$ supergravity in $D=4$, the loop order where a divergence first appears is lowered from $L=7$ to $L=5$~\cite{Bjornsson:2010wm}.  See Refs.~\cite{Bossard:2012xs, Bossard:2013rza, Bern:2013qca, Kallosh:2018wzz, Gunaydin:2018kdz, Carrasco:2022lbm} for various attempts to put tighter restrictions on the counterterms using symmetry, including double-copy consistency, and the associated difficulties.


\vskip -.7 cm $\null$
\subsection{Explicit calculations of counterterm coefficients}
\vskip -.2 cm 

\begin{table}[tb]
\begin{center}
\begin{tabular}{c|c}
 Loops \hskip .2 cm  & \hskip .1 cm critical dimension  \\
 [2pt] \hline \\[-7 pt]
1 & \hskip .1 cm  8 \\
2 &  \hskip .1 cm 7 \\
3 &  \hskip .1 cm 6 \\
4 &  \hskip .1 cm 11/2 \\
5 &  \hskip .1 cm 24/5 \\
\end{tabular}
\vskip .3 cm
\caption{The critical dimensions $D_c$ where ultraviolet divergences first occur in
maximal $\NeqEight$ supergravity, as proven by explicit calculations~\cite{BDDPR,Bern:2007hh,Bern:2008pv, Bern:2009kd,SimplifyingBCJ,UVFiveLoops}.
}
 \label{CriticalDimTable}
\end{center}
\end{table}

A central question is whether the potential counterterms in \tab{GoodCounterTermsTable} lead to actual divergences in the respective theories.  The systematic way to settle this is to carry out direct computations in a way that avoids potential subtleties or unproven arguments. We first describe the status of maximal $\NeqEight$ supergravity in four space-time dimensions before turning to cases with less supersymmetry. As discussed in \sect{LoopLevelSection}, the unitarity method~\cite{UnitarityMethod, Fusing, FiveLoop} in conjunction with the double-copy~\cite{Kawai:1985xq, BCJ,BCJLoop,GeneralizedDoubleCopy} is used to construct amplitudes' integrands, from which the ultraviolet divergences are extracted.  This gives us the coefficient of any potential counterterm, exposing any hidden cancellations~\cite{BDDPR,Bern:2007hh,Bern:2008pv, Bern:2009kd,SimplifyingBCJ,UVFiveLoops}.  String-theory calculations also provide important information~\cite{Green:1982sw, Green:2010sp, Tourkine:2012ip}. Calculations demonstrating the power counting of unitarity cuts~\cite{Bern:2006kd, Bern:2007xj, Herrmann:2018dja, Edison:2019ovj} also provide important insight.  Here we will review the results of such computations, and emphasize various cancellations that are unexplained by symmetry.

In \tab{CriticalDimTable} we collect the critical dimensions where explicit calculations show that maximally supersymmetric supergravity first diverges.  In contrast to expectations in the early 1980s, these calculations show that $\NeqEight$ supergravity is finite though at least five loops.  As described above, these ultraviolet cancellations were subsequently understood to follow from supersymmetry and the $E_{7(7)}$ duality symmetry of $\NeqEight$ supergravity, with the first $D=4$ potential divergence delayed until seven-loop order.

\newcommand{\inlineblob}[1]{\, \vcenter{\hbox{\includegraphics[height=4em]{figs/#1}}}}
\newcommand{\inlinefig}[1]{\,\vcenter{\hbox{\includegraphics[height=3.2em]{figs/#1}}}}
\newcommand{\inlineg}[2]{\inlinefig{Vacuum#1loopsV#2.eps}}
\newcommand{\inlineym}[2]{\inlinefig{Vacuum#1loopsYMV#2.eps}}

To carry out such calculations one first constructs an integrand in ${\cal N}=4$ SYM theory and then promote it through double copy to an ${\cal N}=8$ integrand
\begin{equation}
    (\NeqEight\ \hbox{supergravity}) \sim (\NeqFour\ \hbox{SYM}) \times (\NeqFour\ \hbox{SYM}) \,.
\end{equation}
The integrand obtained in this way is then expanded for large loop momenta or equivalently for small external momenta in order to extract the ultraviolet divergences.\footnote{At five loops this process is more complicated because of difficulties of find five loop integrands that satisfy color-kinematics duality; in this case a generalized double copy procedure is instead applied~\cite{GeneralizedDoubleCopy, GeneralizedDoubleCopyFiveLoops} .}   This expansion gives a set of vacuum-like diagrams that can be reduced to a small number of ``basis integrals." Using integration by parts identities~\cite{Chetyrkin:1981qh}.  The remaining basis integrals are usually simple enough to be analytically integrated giving the exact value of the ultraviolet divergence.  

After reducing the vacuum-like integrals we obtain a simple description of the leading ultraviolet behavior in terms of a basis of vacuum integrals defined as
\begin{equation}
V = -i^{L + \sum_j A_j} \int \prod_{i = 1}^L \frac{d^D \ell_i}{(2 \pi)^D}  \prod_{j} \frac{1}{(p_j^2 -m^2)^{A_j}} \,,
\label{VacuumInts}
\end{equation}
where the $p_i$ are linear combinations of the independent loop momenta and the $A_i$ are the propagators' exponents.
To make the integrals well defined we use dimensional regularization $D = D_c - 2 \epsilon$ where $D_c$ is the critical dimension where the first divergence appears.  In dimensional regularization any divergence will appear as a $1/\epsilon$. It is useful to introduce a mass as an infrared regulator to make it easier to 
separate out infrared singularities from the ultraviolet ones. 

Collecting the results from Refs.~\cite{Green:1982sw, BDDPR, Bern:2007hh, Bern:2008pv, Bern:2009kd, SimplifyingBCJ, UVFiveLoops} the leading ultraviolet behavior of $\NeqEight$ supergravity at each loop order through five loops is described by vacuum diagrams as, where the vacuum integrals are to be evaluated using  dimensional regularization around the critical dimension listed in \tab{CriticalDimTable},
\begin{align}
\null \hskip -.1 cm 
{\cal M}_4^{(1)}\Bigl|_{\rm leading}^{D_c = 8} & = -3  \, \fancyM  \left(\frac{\kappa}{2} \right)^4
 \,\inlinefig{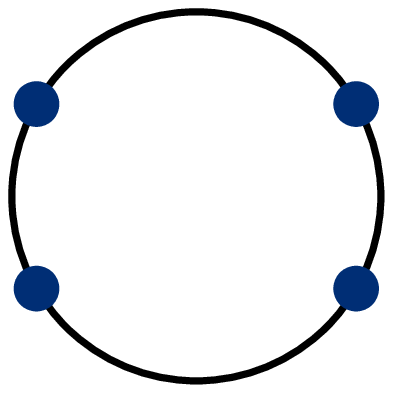} \,,\label{SGVacuum}\\[2pt]
\null \hskip -.1 cm 
{\cal M}_4^{(2)}\Bigl|_{\rm leading}^{D_c = 7}  & = - 8  \, \fancyM  \left(\frac{\kappa}{2} \right)^6
(s^2+t^2+ u^2)
 \left({\ts \frac{1}{4}}  \inlinefig{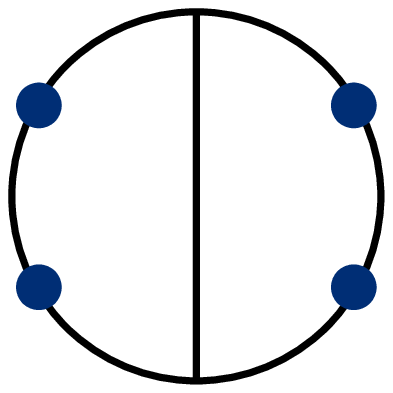}
   + {\ts \frac{1}{4}} \inlinefig{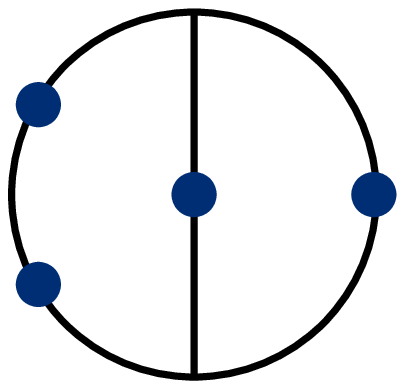} \right) , \nn \\[2pt]
\null \hskip -.1 cm 
{\cal M}_4^{(3)}\Bigr|_{\rm leading}^{D_c = 6}  & =
- 60\,  \fancyM  \left(\frac{\kappa}{2} \right)^8\, s t u 
\left( {\ts \frac{1}{6}}  \inlinefig{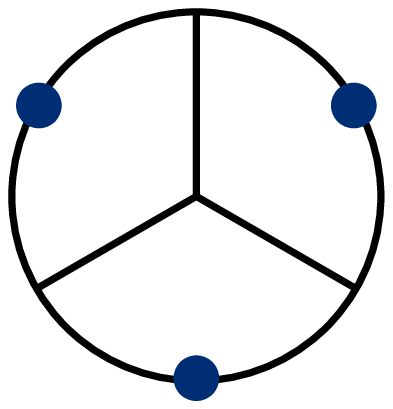} +
 \, {\ts \frac{1}{2}}  \inlinefig{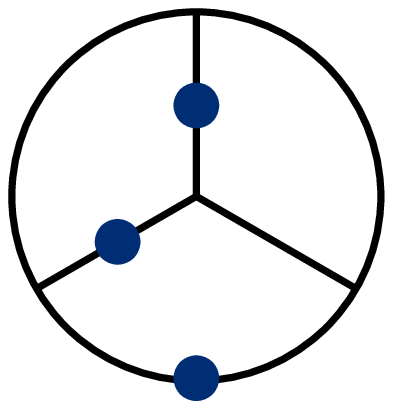} \right) , \nn \\[2pt]
\null \hskip -.1 cm 
{\cal M}_4^{(4)}\Bigr|_{\rm leading}^{D_c = 11/2}  &\hskip -.2 cm  = -\frac{23}{2}  \, \fancyM  
\Bigl(\frac{\kappa}{2}\Bigr)^{10}
 ( s^2 + t^2 + u^2)^2 \left( \! {\ts \frac{1}{4}} \! \inlinefig{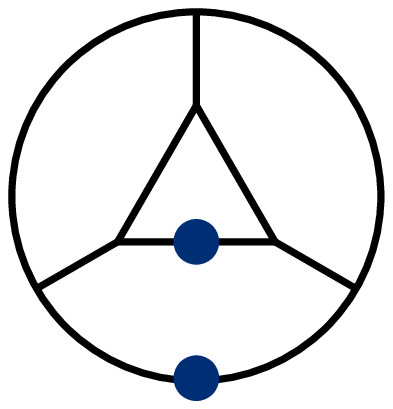}
  + {\ts \frac{1}{2}}\! \inlinefig{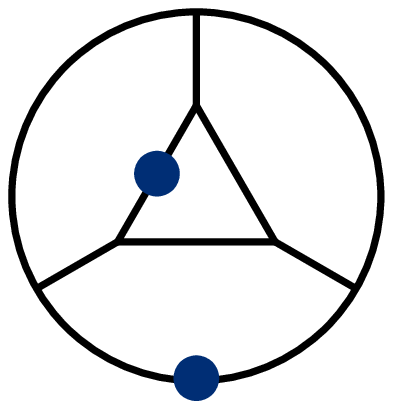}
  + {\ts \frac{1}{4}} \!  \inlinefig{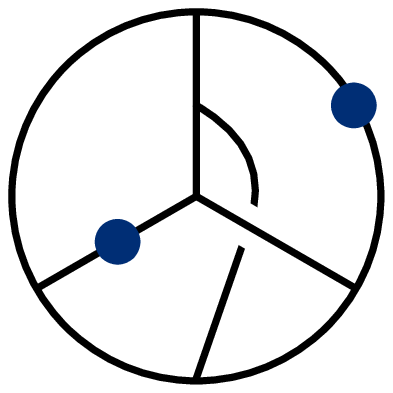}  \right), \nn\\[2pt]
\null \hskip -.1 cm 
{\cal M}_4^{(5)} \Bigr|_{\rm leading}^{D_c = 24/5} &  \hskip -.2 cm =
- \frac{16\times 629}{25} \,  \fancyM  \left(\frac{\kappa}{2}\right)^{12}
(s^2 + t^2 + u^2)^2  
\left({\ts \frac{1}{48}} \inlinefig{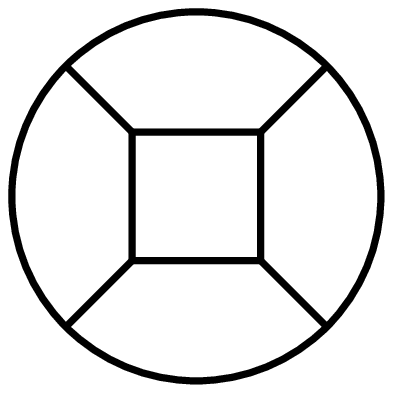}
    + {\ts \frac{1}{16}} \inlinefig{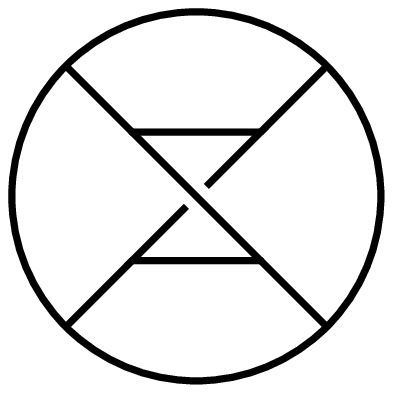}  \right) ,
\nn 
\end{align}
where the prefactor is proportional to the tree amplitude, $\fancyM \equiv s t u \, M_4^\tree(1,2,3,4)$.  The lines represent propagators appearing in \eqn{VacuumInts} and the number of dots $j$ on propagator corresponds to the $A_j -1$ for $A_j \ge 2$ where $A_j$ is the index appearing in \eqn{VacuumInts}. Explicit values of the integrals are found in the source papers, but for our purposes here it is sufficient to note that the basis integrals are all divergent with a definite sign so no further cancellations are possible.

Given the wealth of information contained in \eqn{SGVacuum} one might wonder if there is a path to finding a shortcut to the results, perhaps even to write down the exact divergence in the critical dimension in terms of vacuum integrals to all loop orders. An attempt in this direction is found in Ref.~\cite{UVFiveLoops}.  Unfortunately, the constraints from factorization-type arguments do not appear to be sufficient to uniquely determine the vacuum diagrams at higher orders.  Nevertheless, with additional information it might still be possible to make an educated guess for the exact form of the divergences in $\NeqEight$ supergravity in the critical dimensions. 

If the supersymmetry is reduced the potential counterterms become more complicated, but the loop order where they might first occur is lowered as well.   The case of $N=4$ supergravity is particularly interesting.   As discussed in previous sections via the double pure $\NeqFour$ supergravity can be decomposed into gauge theories,
\begin{equation}
    (\NeqFour\ \hbox{supergravity}) \sim (\hbox{pure YM}) \times (\NeqFour\ \hbox{SYM}) \,.
    \label{Neq4DoubleCopy}
\end{equation}
While it does not have a divergence at the three loops~\cite{Bern:2012cd} contrary to symmetry expectations~\cite{Green:1982sw, Marcus:1984ei, Howe:1988qz},  it, however, does diverge four loops as proven by explicit calculation~\cite{Bern:2013uka}.
The counterterm can be expressed in terms of four powers of the Riemann tensor with the explicit form, 
\begin{align}
C^{L=4, {\cal N} = 4} = -\frac{1}{(4\pi)^8}\left(\frac{\kappa}{2}\right)^6
\frac{1}{72 \epsilon} (1-264\zeta_3)(T_1+2 T_2)\,,
\label{N4Counterterm}
\end{align}
where
\begin{align}
T_1 & \equiv (D_{\alpha}R_{\mu\nu\lambda\gamma})
(D^{\alpha}R_{\rho\sigma}^{\hphantom{\rho\sigma}\lambda\gamma})
R^{\nu\rho}_{\hphantom{\nu\rho}\delta\kappa}
R^{\sigma\mu\delta\kappa}\,, \nn \\
T_2 & \equiv (D_{\alpha}R_{\mu\nu\lambda\gamma})
(D^{\alpha}R_{\rho\sigma}^{\hphantom{\rho\sigma}\lambda\gamma})
R^{\mu\nu}_{\hphantom{\mu\nu}\delta\kappa}
R^{\rho\sigma\delta\kappa}\,.
\end{align}
This expression drops evanescent contributions that vanish in strictly four dimensions and are nonsingular in $D = 4 - 2 \epsilon$ with $\epsilon$ small. Using the helicity form of the divergence calculated in Ref.~\cite{Bern:2013uka} one can also obtain expressions for the counterterms of other states. Indeed, superspace forms of such operators are discussed in Refs.~\cite{Kallosh:1980fi, Howe:1980th, Howe:1981xy, Carrasco:2013ypa}. 

An important question is how  the four-loop divergence of pure $\NeqFour$ supergravity \eqref{N4Counterterm} should be interpreted. A complication is that this theory has a rigid $U(1)$ duality-symmetry anomaly~\cite{MarcusAnomaly, Carrasco:2013ypa}.    Interestingly, the helicity dependence of contributions of the unitarity cuts~\cite{Carrasco:2013ypa} and the divergence~\cite{Bern:2013qca} suggest that the divergence is tied to the $U(1)$ duality anomaly of the theory. Without the anomaly, the $-+++$ and $++++$ helicity sectors would vanish. Ref.~\cite{Carrasco:2013ypa} demonstrates that the anomaly leads to a poor ultraviolet behavior in the $--++$ as well.  In this way the divergence is directly connected to the appearance of anomalous amplitudes which vanish at tree level but are nonzero starting at one loop~\cite{Cangemi:1996rx}.  It would be important to demonstrate this directly, presumably by tracking the anomaly contributions to four-loop amplitudes and showing that all but anomaly terms cancel. One may also wonder whether it is possible to remove the divergence by adding a finite term to the action so that an appropriate symmetry required for finiteness is preserved~\cite{Bern:2017rjw}.  This brings up the question of whether the pure supergravity theories, which do not have such anomalies, are ultraviolet finite, or at least have an improved behavior~\cite{Bern:2013uka, Bern:2017rjw, Bern:2019isl}. 

In $\NeqFive$ supergravity in $D=4$ there are no anomalous amplitudes of the type that appear in $\NeqFour$ supergravity~\cite{Freedman:2017zgq} making it an ideal case to investigate.   From the double-copy perspective this theory is given as a product of 
\begin{equation}
(\NeqFive\ \hbox{supergravity}) \sim (\NeqOne\  \hbox{SYM}) \times (\NeqFour\ \hbox{SYM}) \,.
\end{equation}
The computation of the potential four-loop divergence in this theory follows closely the one of in $\NeqFour$ supergravity in four dimensions.   In this case, the potential counterterm vanishes~\cite{Bern:2014sna}
\begin{equation}
C^{L=4, {\cal N} = 5} = 0\,. 
\end{equation}
This vanishing is nontrivially involving cancellations between planar and nonplanar contributions.  The lack of a robust symmetry argument for this vanishing has been confirmed in Ref.~\cite{Freedman:2018mrv}.

A particularly striking example of a theory exhibiting enhanced cancellations is pure half-maximal supergravity in $D=5$.  The first potential counterterm is a two loops, corresponding to an $R^4$ counterterm. The double-copy construction for this case is
\begin{equation}
    (\hbox{half-maximal supergravity}) \sim (\hbox{pure YM}) \times (\hbox{maximal SYM}) \,,
\end{equation}
which is essentially the same one as for $\NeqFour$ supergravity in $D=4$ \eqref{Neq4DoubleCopy}, except that the two component gauge theories are promoted to $D=5$.   This case is simple enough so that the cancellations can be analyzed in detail~\cite{Bern:2012gh}.  By assuming the existence of an off-shell superspace that manifests all supersymmetries such a counterterm can be ruled out~\cite{Bossard:2013rza}; however, doubt is cast on the existence of such a superspace by it also predicting finiteness for certain matter-coupled supergravities for which explicit divergences were found~\cite{Bern:2013qca}. 
 
Despite the lack of a symmetry argument, the coefficient of the potential $L=2$, $D=5$ divergence in pure half-maximal supergravity does indeed vanish~\cite{Bern:2012gh}. It is noteworthy that the absence of divergences~\cite{Tourkine:2012ip} in half-maximal supergravity in $D = 5$ has also been obtained from string-theory calculations, offering an alternative approach to expose the ultraviolet cancellations in this theory. 

Remarkably, the enhanced cancellations for this example are directly tied to the double-copy structure inherent in the amplitudes~\cite{Bern:2012gh}.  The diagrams that result from the double copy are individually ultraviolet divergent with quadratic divergences appearing at two loops.  Nevertheless, when all diagrams are combined the divergences cancel from the amplitude~\cite{Bern:2012gh}. This case is particularly simple to analyze because of special properties of the two-loop maximal SYM four- and five-point amplitudes.  These amplitudes carry no powers of loop momentum in the kinematic numerators, making it straightforward to directly express the supergravity amplitudes after loop integration directly in terms of integrated pure-Yang--Mills amplitudes.  This allows us to directly relate two-loop four- and five-point potential divergences in supergravity to corresponding ones of pure Yang--Mills theory. Gauge invariance dictates the form of the potential counterterms in Yang--Mills theory, putting restrictions on the color factors that can appear.  Through the double copy, restrictions on gauge-theory divergences then imply restrictions on the supergravity divergences.  In this way finiteness of half-maximal supergravity four- and five-point amplitudes at one loop in $D < 8$ and at two loops in $D < 6$ can be understood, with the cancellations being looked to those of well understood forbidden color factors in corresponding pure Yang--Mills amplitudes.  Unfortunately, it is not straightforward to extend this analysis to higher loops because the integrals one encounters are no longer identical to the ones encountered in gauge theory. Nevertheless, this does demonstrate the importance of hidden structures of symmetries severely limiting the form of supergravity divergences.

\section{Web of supergravity theories}
\label{WebSection}

\def\no{\nonumber}
\def\ha{\hat a}
\def\hb{\hat b}
\def\hc{\hat c}
\def\hd{\hat d}
\def\he{\hat e}
\def\hA{\hat A}
\def\hB{\hat B}
\def\hC{\hat C}
\def\hD{\hat D}
\def\hE{\hat E}

\def\tR{t_{\cal R}}

\newcommand{\co}{\ , \ \ \ \ \ \ }
\newcommand{\dd}{\mathrm{d}}
\newcommand{\te}{\textrm}
\newcommand{\al}{\alpha}
\newcommand{\la}{\lambda}
\newcommand{\vph}{\varphi}
\newcommand{\ap}{{\alpha'}}

\newcommand{\haa}{{\hat \alpha}}
\newcommand{\hbb}{{\hat \beta}}
\newcommand{\hgg}{{\hat \gamma}}
\newcommand{\hdd}{{\hat \delta}}
\newcommand{\hee}{{\hat \epsilon}}

\def\cN{{\cal N}}
\def\wone{0.3\textwidth}
\def\wtwo{0.50\textwidth}
\def\wfour{0.14\textwidth}
\def\wthree{0.6\textwidth}
\def\wfive{0.4\textwidth}

\setlength{\LTcapwidth}{\textwidth}

\begin{figure}[t]
\begin{center}
  \includegraphics[width=.9\textwidth]{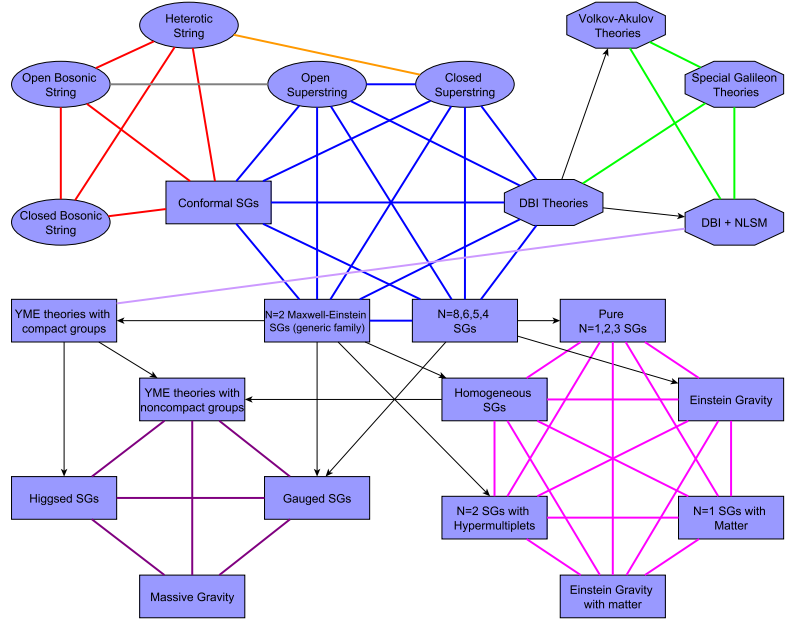} 
 \medskip 
\caption{Web of double-copy-constructible theories.
 Links with different colors connect theories that have a
 common gauge theory in their construction. 
 Directed links point toward constructions that are obtained by modifying both gauge-theory
 factors (see Refs.  \cite{Bern:2019prr,Bern:2022wqg}). The main text provides a discussion of theories on/close to the central hexagon.   }
\label{FigWeb}
\end{center}
\end{figure}

Up to this point, we have studied the double-copy structure and ultraviolet properties mostly within the context of $\mathcal N \ge 4$ supergravities.  The reader may wonder whether similar progress can be made for other classes of gravity theories. 

As we shall briefly discuss in this section, there is by now extensive evidence that very large families of gravitational theories can be seen as double copies. However, at present, it is not clear whether all gravities are double-copy constructible and, if not, what are the criteria that determine whether a double copy can be formulated for a given theory. In figure \ref{FigWeb}, we give a pictorial rendition of the web of theories linked by common gauge-theory factors in their double-copy constructions. While there is not sufficient space here to discuss all individual nodes, we briefly comment on some examples.   For further details,  the reader is invited to study  the reviews~\cite{Bern:2019prr} and \cite{Bern:2022wqg}.

\vskip -.7 cm $\null$
\subsection{Purely adjoint double copies}
\vskip -.2 cm 

The fundamental ingredient entering a double-copy construction is the set of duality-satisfying numerators from a gauge theory. We have also seen that both SYM theory with various amounts of supersymmetry and pure Yang--Mills theory in any dimension satisfy the duality. In addition to that, if we restrict our analysis to gauge theories with only adjoint fields, it is relatively straightforward to find additional simple examples of theories that obey color-kinematics duality:
\begin{itemize}
\item One can consider consistent truncations (sometimes called field-theory orbifolds) of $\cN=4$ SYM theory that preserve a portion of the original supersymmetry and only have adjoint fields (see Refs. \cite{Carrasco:2012ca, Chiodaroli2013upa, Damgaard2012fb} for a comprehensive discussion). They yield, aside from pure $\cN=1,2$ SYM theories, also $\cN=1,2$ SYM with one vector multiplet. 
\item The dimensional reduction of a pure Yang--Mills theory in $D$-dimensions to $d$ dimensions gives a Yang--Mills-scalar theory with adjoint scalars and a $SO(D-d)$ global symmetry. Note that, in general, dimensional reduction preserves color-kinematics duality. 
\item The latter theory can be deformed introducing trilinear scalar couplings of the form \cite{Chiodaroli2014xia},
\begin{equation}
\delta {\cal L} =
\frac{\lambda} {6!} F^{IJK}  {\rm Tr} [\phi^I, \phi^J ] \phi^K \,, \label{tri-couplings}
\end{equation}
where $F^{IJK}$ are structure constants for a  subgroup of the $SO(D-d)$ global symmetry, $I,J,K$ are global indices carried by the scalars, and $\lambda$ is a free parameter. In this case, color-kinematics duality implies that the $F^{IJK}$ tensors obey Jacobi relations.
\item The nonlinear sigma model (NLSM) can also be shown to obey color-kinematics duality, with the caveat that in this case the color gauge group is replaced by a global (flavor) group~\cite{Chen2013fya, Cachazo2014xea, Chen2014dfa, Du2016tbc, Chen2016zwe, Carrasco2016ygv, Cheung:2016prv,Pavao:2022kog}. The NLSM is a sufficiently-simple example that a Lagrangian manifesting the duality is available, see Ref.~\cite{Cheung:2016prv}.  
\end{itemize}
Aside from the above cases,  there is
another known gauge theory with only adjoint fields that obeys color-kinematics duality: a higher-derivative version of Yang--Mills theory with a mass parameter that has been dubbed $(DF)^2$ theory~\cite{Johansson:2017srf,JohanssonConformal,Carrasco:2022lbm}. While there exist several distinct versions, the simplest (minimal) theory has Lagrangian,
\vskip -.5 cm 
\begin{align}
	{\cal L}_{(DF)^2+{\rm YM}}&= \frac{1}{2}(D_{\mu} F^{a\, \mu \nu})^2- \frac{1}{4} m^2 (F^a_{\mu \nu})^2 \,,
\end{align}
where $a$ is an adjoint index. There also exist variants of the theory with additional scalars in a specific matter representation \cite{Johansson:2017srf,JohanssonConformal,Carrasco:2022sck}.

\def\wdone{0.6\textwidth}
\def\wdzero{0.3\textwidth}

\begin{table}[tb]
\begin{center}
\begin{tabular}{c@{}|@{\ \ \ \ }c}
  \backslashbox{\ \ \ \ \ \ QFT \ \ \ \ \ \ \ \ \ \ \ \  }{\! \! \! QFT \ \ \ \ \ \ \ } &  
 $\cN=N_1$ SYM theory  \\[1pt]
 \hline
 \\[-5pt]
 $\cN=N_2$ SYM \ \ & 
 $\cN=N_1+N_2$ 
 Maxwell-Einstein 
 Supergravity 
   \\[5pt] 
 NLSM \ \ & 
 $\cN=N_1$ 
 Dirac-Born-Infeld theory     \\[5pt]
 YM-scalar  from dim. red.  & 
 $\cN=N_1$ 
 Supergravity with 
 matter vector multiplets   \\[5pt]
 YM + $\phi^3$  theory  & 
 $\cN=N_1$ 
 Yang--Mills-Einstein  Supergravity 
  \\[5pt]
$(DF)^2$ 
 theory  & 
$\cN=N_1$ 
Conformal Supergravity 
  \\[5pt]
\end{tabular}
\medskip
\caption{Multiplication table for purely-adjoint double copies involving at least one SYM gauge theory with $\mathcal N = {\mathcal N}_1$ supersymmetry. The left column gives the second theory used in the double-copy construction.  Theories are specified in $D$ dimension. In  case of the Yang--Mills-scalar theory from dimensional reduction, the theory is reduced  from higher dimensions. See also Fig. \ref{FigWeb}.}
\label{SYMdoublecopies}
\end{center}
\vskip -.5 cm 
\end{table}

To give concrete examples, the right column of Table~\ref{SYMdoublecopies} gives a variety of double-copy theories where one copy is $\mathcal N = {\mathcal N}_1$ SYM theory with the second copy shown in the left column.   These results are located on/close to the central hexagon of Fig.~\ref{FigWeb} with blue links representing SYM theory.

The first row corresponding to the previously explained the double copy of two SYM theories leads to an ungauged supergravities with $N=N_1 +N_2$ supersymmetries. With $D=4$, for $\cN\leq 4$, the result is supergravities including matter multiplets; specifically, if $N_1=N_2=2$ we obtain ungauged $\cN=4$ supergravity in four dimensions with two vector multiplets, if $N_1=N_2=1$, we obtain $\cN=2$ supergravity in four dimensions with one hypermultiplet~\cite{BCJ, BCJLoop ,N46Sugra, Bern:2009kd, Bern:2014sna, Carrasco:2012ca}.  A simple way to understand the appearance of different multiplets is by counting the number of resulting states in the double-copy tensor products of states.

The second row in  Table~\ref{SYMdoublecopies} gives the double-copy between the NLSM and SYM theory, which produces amplitudes of a supersymmetric version of the Dirac-Born-Infeld theory~\cite{Cachazo2014xea,Cheung:2017ems}. This is not actually a gravitational theory: since the NLSM does not contain gluons, a graviton cannot be obtained from the double copy.

An important class of double-copy constructions arises when one of the gauge-theory factors is non-supersymmetric, as in the third row of  Table~\ref{SYMdoublecopies}. In this case, one has more freedom.
The simplest option is the dimensional reduction of a pure Yang--Mills theory from $n+4$ or $n+5$ dimensions to four or five dimension. The double copy with SYM theory leads in this case---according to the amount of supersymmetry---to either a $\cN=4$ ungauged supergravity with $n$ vector multiplets, or to a $\cN=2$ Maxwell-Einstein supergravity belonging to the so-called generic Jordan family~\cite{Chiodaroli2014xia} (see Refs. \cite{Gunaydin:1983bi,Gunaydin1984ak} for a supergravity discussion of these theories).

An interesting variant to the latter construction is obtained by adding trilinear scalar coupling in \eqn{tri-couplings}~\cite{Chiodaroli2014xia}, as it appears on the fourth row of  Table~\ref{SYMdoublecopies}; the net result is either a $\cN=2$ or a $\cN=4$ Yang--Mills-Einstein supergravity in which the $F^{IJK}$-tensors from (\ref{tri-couplings}) are identified with the structure constants of the supergravity gauge group (see also Refs. \cite{Cachazo2014nsa, Chiodaroli:2016jqw, Chiodaroli:2017ngp}). Note that the R-symmetry in untouched by this construction. 

Finally, the minimal $(DF)^2$  theory enters the last row of Table~\ref{SYMdoublecopies}. Together with SYM theory, its double-copy construction gives a mass deformation of conformal supergravity~\cite{Johansson:2017srf, JohanssonConformal, Menezes:2021dyp},
\begin{equation}
\big( \text{mass-deformed minimal CSG} \big)=\big({\rm SYM}\big) \otimes \big(\textrm{minimal  $(DF)^2$ + YM}\big)\, .
\label{CG_DC_massdef}
\end{equation}
The constructions outlined so far can be implemented at tree level, both by combining sets of duality-satisfying numerators as in Eq.~\ref{squaring}, or by using the KLT formula \ref{KLT45}. 

It is interesting to note that these type of double-copy constructions extend well outside the realm of ordinary supergravity theories. In particular, with a slightly different construction, we can obtain amplitudes from string theory. The starting point for this family of constructions is given by the set of  disk integrals~\cite{MafraBCJAmplString,Broedel2013tta},
\begin{equation}
Z_\sigma(\rho(1,2,\ldots,n)) = (2 \ap)^{n-3} \! \! \! \! \! \! \! \! \! \! \! \! \! \! \! \! \! \! \! \!  \! \! \! \!\int \limits_{\sigma \, \{  - \infty \leq z_{1}\leq z_{2} \leq \ldots \leq z_{n}\leq \infty \}} \! \! \! \! \! \! \! \! \! \! \! \! \! \! \! \! \! \! \! \! \! \frac{d z_1 \, \ldots\, d z_{n}}{{\rm vol}({\rm SL}(2,\mathbb{R}))} \ 
\frac{ \prod_{i<j}^{n} |z_{ij}|^{\alpha' s_{ij}}  }{ \rho \, \{ z_{12} z_{23} \cdots z_{n-1,n} z_{n,1} \} } \, .
\label{discint}
\end{equation}
In this equation $z_{ij}=z_i-z_j$ indicates the difference between the positions of the punctures, The  ${\rm vol}({\rm SL}(2,\mathbb R)$ factor is handled by fixing the position of three punctures while introducing an appropriate Jacobian. The above integrals  satisfy~\cite{Broedel2013tta} the field-theory fundamental  BCJ relations (\ref{BCJrels}) with respect to the permutation~$\rho$. Hence, they can be naturally contracted by the field-theory KLT kernel to give a well defined double copy. 

Taking SYM theory for the other set of partial ordered amplitudes, it can be shown that the output of the double-copy results in open-superstring amplitudes with color-ordered massless external states ~\cite{MafraBCJAmplString,Broedel2013tta},
\begin{equation}
\hskip -.2 cm 
A^{\tree}_{\rm OS}(\sigma(1,2,\ldots,n)) ~ = \!\!\!\!\!\!\!\!\!\!\!  \sum_{\tau,\rho \in S_{n-3}(2,...,n-2)}\!\!\!\!\!\!\!\!\!\!\!  Z_\sigma (1,\tau,n,n{-}1)  S[\tau | \rho]  A_{\rm SYM}(1,\rho,n{-}1,n) \,. \hskip .1 cm 
\label{2.2cOLD} 
\end{equation}
The $Z$ integrals have been interpreted as
the amplitudes of a scalar theory dubbed Z-theory in Refs.~\cite{Carrasco2016ldy, Mafra2016mcc, Carrasco2016ygv}.   A closed-string version of the Z-theory integrals is also available~\cite{Stieberger2014hba, Schlotterer2018zce, Vanhove:2018elu, Brown:2018omk},
\begin{align}
{\rm sv} \, Z(\tau  | \sigma) & = \left( \frac{2\ap}{\pi } \right)^{n-3}  \! \! \!  \int \!\!  \frac{d^2 z_1  \ldots d^2 z_{n}}{{\rm vol}({\rm SL}(2,\mathbb{C}))} \,
\frac{ \prod_{i<j}^{n} |z_{ij}|^{2 \alpha' s_{ij}}  }{ \tau \, \{ \bar{z}_{12}  \cdots \bar{z}_{n-1,n} \bar{z}_{n,1} \}  \sigma \, \{ z_{12}  \cdots z_{n-1,n} z_{n,1} \} } \,. \nonumber \\
& \label{2.2bcl}
\end{align}
Here we consider ``sv" as part of the name of these integrals, but it can also be understood as an operation known as single-value projection (see Refs.~\cite{Schnetz2013hqa, Brown2013gia}).
Using these building blocks, closed-superstring amplitudes can be constructed as double copies \cite{Schlotterer:2012ny,Stieberger:2013wea}
\be
(\text{closed superstring})= ({\rm SYM}) \otimes  {\rm sv}\big(\textrm{open superstring} \big) \, .
\label{CSvsOS}
\ee
This can also be understood as a ``triple copy"
\be
(\text{closed superstring})= ({\rm SYM}) \otimes  {\rm sv}\big(\textrm{Z-theory} \big) \otimes ({\rm SYM}) \, .
\label{CSvsOS2}
\ee
Morevoer, an analogous construction has  been formulated for the open bosonic string \cite{Azevedo:2018dgo}
\be
(\text{open bosonic string})= ({(DF)^2 \rm  theory}) \otimes  \big(\textrm{Z-theory} \big) \, .
\label{CSvsOBS}
\ee
Relevant stringy double-copy constructions are summarized in Table \ref{stringydoublecopies}.

\def\wdone{0.6\textwidth}
\def\wdzero{0.3\textwidth}

\begin{table}[tb]
\begin{center}
\begin{tabular}{c|@{\ \ \ \ \ \ \ \ \ \ }c}
 \backslashbox{\ \ \ \ \ String  \ \ \ \ \ \ \ \ \  }{\! \! \! QFT \ \ \ \ \ \ \ } &  SYM theory  \\[5pt]
 \hline 
 Z-theory \ \ & 
  Open superstring  \\[5pt] 
 sv(open superstring) \ \ & 
 Closed superstring \\[5pt]  
  sv(open bosonic string)  & 
 Heterotic string (gravity sector) 
 \\  
\end{tabular}
\vskip .3 cm
\caption{Multiplication table for stringy double copies involving at least one SYM~\cite{Schlotterer:2012ny, Stieberger:2013wea, Azevedo:2018dgo}. }
\label{stringydoublecopies}
\end{center}
\vskip -.5 cm 
\end{table}

Many additional variants of the construction become available if one includes gauge theories with extra fields transforming in matter representations on top of the adjoint sector.  In order to describe this family of constructions, one needs to spell out some additional rules on how fields in different representations enter the double-copy construction. 

A common starting point is to consider gauge theories whose amplitudes can still be obtained in a presentation based on cubic graphs, i.e. there is no quartic or higher invariant symbol for the gauge group, so that only representation matrices, structure constants and possibly Clebsch-Gordan coefficients with three indices from different representations appear in the color-dressed amplitudes. In this case, a color-kinematics-duality-satisfying presentation of amplitudes can still be obtained if numerator factors obey the same algebraic relations as the color factors, which now also include commutation relations for the representation matrices \cite{Chiodaroli:2015rdg, Bern:2019prr}.

In order to combine fields in matter representations to construct double-copy states, we associate a gravitational state in the output of the construction to each gauge-invariant bilinear built out of gauge theory. In practice, this implies that no double-copy state is constructed as the product of an adjoint and a matter-representation state. The matter representations simply yield additional sectors in the supergravity theory. This criterion allows us to avoid the appearance of the extra spin-$3/2$ states that would otherwise lead to inconsistencies as discussed in Sect.~\ref{sec:ckdAndSusy}.  

A concrete example of this family of double-copy constructions is given by the so-called homogeneous $\cN=2$ supergravities in four and five dimensions. These theories have been first classified from the supergravity perspective by de~Wit and van~Proeyen in Ref.~\cite{deWit1991nm}, and possess scalar manifolds that are usually referred to as $L(q,P)$ spaces. Their double-copy construction was formulated more recently in Ref.~\cite{Chiodaroli2015wal}, and have the structure
\begin{align}
\left(
\begin{array}{c}
\cN  =   2  \text{ homogeneous} \\
\text{ supergravity }
\end{array} \right)
 \; \; = \; \;  
\left(
\begin{array}{c}
 \cN = 2 \text{ SYM}  \\
 \null + \frac{1}{2} \text{ hyper}_R
\end{array} \right) 
 \otimes
\left(
\begin{array}{c}
\text{YM} \text{ from D dimensions}  \\ \null + n_f \text{ fermions}_R
\end{array} \right)  \,. \label{DChom}
\end{align}
The supersymmetric gauge theory is $\cN=2$ SYM with a half-hypermultiplet transforming in a matter pseudo-real representation.\footnote{Pseudo-reality of the representation is a necessary ingredient in the construction and ensures that the half-hyper can be introduced in the theory without having to be completed to a full hyper, see Ref.~\cite{Chiodaroli2015wal}.}
The non-supersymmetric theory entering the construction is a Yang--Mills theory with $n_f$ irreducible fermions in $D$ dimensions reduced to four or five dimensions. The number of matter fermions and the dimension for  the non-supersymmetric theory are related to the parameters given in the supergravity literature as $D=q+6$ and $P=n_f$. Additionally, in some particular dimensions corresponding to $q=0$ (mod $4$), there exist two inequivalent spinor representation with different chirality. This corresponds to an additional parameter in the supergravity construction.  The reader should consult Refs.~\cite{Chiodaroli2015wal,Ben-Shahar:2018uie} for details. 

Another important example is the double-copy construction for some gauged supergravities with Minkowski vacua~\cite{Chiodaroli:2017ehv,Chiodaroli:2018dbu}. These theories are obtained as the double copy of a spontaneously-broken gauge theory with an explicitly broken theory with massive fermions, after carefully tuning gauge-group representations and masses. While this family of constructions is still in its infancy, it provides an example of the subtleties one faces in extending the double copy to theories with several distinct matter representations.   

The reader may wonder how is it possible to provide a double-copy construction for Einstein gravity. This requires removal of unwanted states produced by the double copy, which can be done by introducing matter fermions in one of the two Yang--Mills theories and matter ghost fields in the other~\cite{Johansson2014zca,Luna2017dtq} or by explicit projection~\cite{Carrasco:2021bmu}.

Many additional examples of double-copy constructions have been formulated over the years beyond the ones mentioned in this section. These include: theories with  hypermultiplets~\cite{Chiodaroli2015wal,Anastasiou:2017nsz}, Yang--Mills--Einstein theories with spontaneously-broken gauge symmetry~\cite{Chiodaroli:2015rdg}, various gravitational theories with higher-dimension operators \cite{Broedel2012rc,Carrasco:2019yyn,Carrasco:2021ptp,Chi:2021mio,Carrasco:2022lbm}, gravity with massive matter~\cite{Plefka:2019wyg,Carrasco:2020ywq,Carrasco:2023vjg}, massive (Kaluza-Klein) gravity~\cite{Chiodaroli:2015rdg,Momeni:2020vvr,Momeni:2020hmc,Johnson:2020pny}, heavy-mass effective theories~\cite{Haddad:2020tvs,Brandhuber:2021kpo}, 
and special constructions for supergravities in three dimensions\cite{Huang:2012wr,Bargheer:2012gv,Sivaramakrishnan:2014bpa}.
While here we have focused on  BCJ double copies, similar results have been obtained with the so-called CHY formalism~\cite{Cachazo:2013hca,Cachazo:2013iea, Cachazo2014xea}, as well as with ambitwistor strings~\cite{Mason:2013sva, Adamo:2013tsa, Geyer:2014fka, Geyer:2015jch, Casali2015vta, Geyer:2015bja, Azevedo:2017lkz, Geyer:2017ela, Geyer:2018xwu}.  (See also the review~\cite{Geyer:2022cey}), the KLT bootstrap \cite{Elvang:2020lue}, formalisms based on differential operators~\cite{Cheung:2016drk,Cheung:2016prv, Cheung:2017yef} and approaches based on off-shell linearized (super)fields~\cite{Borsten2013bp, Anastasiou:2013hba, Anastasiou2014qba, Nagy:2014jza, Anastasiou2016csv, Anastasiou:2018rdx}.   

%

\section{Conclusions and Outlook}
\label{ConclusionSection}

This chapter presented an overview of the on-shell perspective 
for supergravity theories~\cite{UnitarityMethod, Fusing, FiveLoop}, focusing on their ultraviolet properties and on the double-copy structure~\cite{Kawai:1985xq, BCJ,BCJLoop, Bern:2019prr}. The double copy offers a radically different interpretation of perturbative gravity, and relates supergravity and gauge-theory calculations at high loop orders, providing practical means to determine explicitly the coefficient of potential ultraviolet counterterms.

The question of ultraviolet divergences in supergravity theories has a long history. The dimensionful nature of Newton's constant suggests that all supergravity theories should diverge at some loop order. By their very nature, such power counting arguments can fail if symmetries or structures that induce nontrivial cancellations are not taken into account. 
This is not a hypothetical issue:  at present there is no explanation for such {enhanced cancellations}. Examples include the absence of an ultraviolet divergence in $\NeqFive$ supergravity at four loops~\cite{Bern:2014sna}, despite the existence of a counterterm that respects all symmetries manifestly preserved by regularized 
scattering amplitudes~\cite{Bossard:2011tq, Freedman:2018mrv, Kallosh:1980fi, Howe:1980th}. Half-maximal supergravity at two loops in $D=5$ is another such example~\cite{Bern:2012gh, Tourkine:2012ip, Bossard:2013rza, Bern:2013qca}. 
An important characteristic of enhanced cancellations is that no local representation of the amplitude in terms of covariant diagrams can display them: the power counting of each diagram is necessarily worse than the power counting of the complete amplitude.  
In other words, the cancellations are highly nontrivial, and require an interplay between  most, if not all, diagrams contributing to an amplitude, thus pointing to some novel structures or symmetries. 

The existence of enhanced cancellations suggests that further nontrivial surprises await us as we probe supergravity theories to ever higher loop orders.
A primary goal is to fully understand the origin of enhanced cancellations. For the case of half-maximal supergravity at two loops in $D=5$, the double copy links the potential ultraviolet divergences to certain would-be divergences in pure-Yang--Mills with forbidden color factors~\cite{Bern:2012gh}. 

The enhanced cancellations that appear in $\NeqFive$ supergravity at four loops in $D=4$ is an important case because the amplitudes are free of the $U(1)$ anomalies that appear in $\NeqFour$ supergravity~\cite{MarcusAnomaly, Carrasco:2013ypa, Bern:2013qca}, which plausibly are an underlying source of divergences. 
An important challenge is to calculate the coefficient of the potential five-loop divergence in $\NeqFive$ supergravity and ascertain whether enhanced cancellations continue beyond four loops.  
The generalized double copy~\cite{GeneralizedDoubleCopy, GeneralizedDoubleCopyFiveLoops} provides a means to carry out this calculation.
If this turns out to be finite, it should invigorate a search for an all-orders proofs of ultraviolet finiteness, analogous to the way calculations through three loops~\cite{Jones:1977zr, Poggio:1977ma, Grisaru:1979wc,Grisaru:1980nk} stimulated proofs of finiteness for $\NeqFour$ SYM theory~\cite{MandelstamN4SYM, BrinkN4SYM, HoweStellN4SYM}. On the other hand, if it turns out to be divergent at five loops, it would go a long way towards finally settling the question of whether ultraviolet-finite supergravity theories exist.
Another important question relates to identifying a complete explanation for enhanced cancellations.  Possible paths include an improved understanding of the consequences of supersymmetry, duality symmetries, or perhaps generalized symmetries \cite{Gaiotto:2014kfa,  Komargodski:2020mxz, Cordova:2022ruw}. Whatever explanation is ultimately found should be novel, given that the enhanced ultraviolet cancellations themselves are not of a type usually found in gauge theories. 

There are also challenges related to improving our understanding of the duality between color and kinematics and the associated double copy. Our ability to carry out high-loop computations relies on finding double-copy format of gravitational scattering amplitudes. In particular, no BCJ representation of the five-loop four-point $\NeqFour$ SYM amplitude is known, making it necessary to use the more complicated generalized double-copy construction~\cite{GeneralizedDoubleCopy,GeneralizedDoubleCopyFiveLoops,UVFiveLoops}.  While this is sufficient for carrying out five-loop computations for $\NeqEight$ supergravity~\cite{GeneralizedDoubleCopyFiveLoops}, and likely for $\NeqFive$ supergravity, such computations would be enormously simplified by having gauge-theory integrands that manifestly satisfy the duality between color and kinematics.  

One of the most interesting aspects of the double copy is that it gives nontrivial links between theories.  This web of theories~\cite{Cheung:2017ems, Carrasco:2019qwr, Bern:2019prr} links together both gravitational and nongravitational theories by sharing single-copy theories. 
The complete picture is still elusive, and may ultimately answer the outstanding question of whether all supergravity theories be expressed in a double-copy form~\cite{Chiodaroli2015wal, Anastasiou:2017nsz}.

Another important outstanding question is to fully understand the underlying kinematic algebra~\cite{Monteiro:2011pc, OConnellAlgebras, Monteiro:2013rya,Fu:2016plh, Fu:2018hpu,Cheung:2016prv, Chen:2019ywi, Chen:2021chy, Cheung:2021zvb, Brandhuber:2021bsf, Cheung:2022mix,Ben-Shahar:2021doh,Ben-Shahar:2021zww}, which would be helpful for building multiloop supergravity integrands through the double copy.  Further interesting related topics which we did not cover here include the construction of classical double-copy solutions, applications to gravitational-wave physics, and generalizations color-kinematics duality and double copy to massive theories and to theories with different representations of the gauge group. For further discussion we refer the reader to various reviews~\cite{Carrasco:2015iwa, Cheung:2017pzi, White:2017mwc, Bern:2019prr, Borsten:2020bgv, Bern:2022wqg, Adamo:2022dcm, Bern:2022jnl}.

In summary, while many interesting questions remain, the double-copy approach to (super)gravity has proven to be rather fruitful, such as finding nontrivial relations to gauge theories, providing a powerful route to high-loop calculations shedding light on the ultraviolet behavior of supergravity theories.

\vskip -.8 cm $\null$
\subsection*{Acknowledgments}
\vskip -.3 cm $\null$
We thank  J. Gates, M. G\"{u}naydin and R. Kallosh for comments on the draft and helpful discussions. 
Z.B. is supported by the U.S. Department of Energy (DOE) under grant no.~DE-SC0009937 and by the Mani L. Bhaumik Institute for Theoretical Physics.  J.J.M.C. is grateful for the support of Northwestern University, the DOE under contract DE-SC0015910, and by the Alfred P. Sloan Foundation. R.R.~is supported by the U.S. Department of Energy (DOE) under grant no.~DE-SC00019066. The work of M.C. is supported by the Swedish Research Council under grant 2019-05283. The research of M.C.~and~H.J. is also supported by the Knut and Alice Wallenberg Foundation under grants KAW 2018.0116 ({\it From Scattering Amplitudes to Gravitational Waves}) and KAW 2018.0162.


\bibliographystyle{iopart-num}

\bibliography{supergravity}

\end{document}